\documentclass[preprint,amsmath,floatfix,prb,aps]{revtex4}

\usepackage{color}
\usepackage{lipsum,amsmath}
\usepackage{pifont}   
\usepackage{graphicx} 
\usepackage{dcolumn}  
\usepackage{bm}       
\usepackage{amsfonts} 
\usepackage{amssymb}  
\usepackage{multirow} 
\usepackage{siunitx}

\newcommand{\BB}{{\bf b}}

\newcommand{\br}{{\bf r}}
\newcommand{\bp}{{\bf p}}
\newcommand{\bk}{{\bf k}}
\newcommand{\bv}{{\bf v}}

\newcommand{\bu}{{\bf u}}
\newcommand{\bq}{{\bf q}}
\newcommand{\bG}{{\bf G}}

\newcommand{\bK}{{\bf K}}

\newcommand{\bA}{{\bf A}}

\newcommand{\ket}[1]{\left| #1 \right\rangle}

\newcommand{\mel}[3]{\left\langle #1 \left| #2 \right| #3 \right\rangle}

\newcommand{\bdel}{\boldsymbol{\delta}}
\newcommand{\bsig}{\boldsymbol{\sigma}}

\newcommand{\bnu}{\boldsymbol{\nu}}

\begin{document}


\title{Perfect and controllable nesting in the small angle twist bilayer graphene}

\author{Maximilian Fleischmann$^1$}
\author{Reena Gupta$^1$}
\author{Florian Wullschl\"ager$^2$}
\author{Dominik Weckbecker$^1$}
\author{Velimir Meded$^3$}
\author{Sangeeta Sharma$^4$}
\author{Bernd Meyer$^2$}
\author{Sam Shallcross$^1$}
\email{sam.shallcross@fau.de}
\affiliation{1 Lehrstuhl f\"ur Theoretische Festk\"orperphysik,
  Friedrich-Alexander-Universit\"at Erlangen-N\"urnberg (FAU),
  Staudtstra{\ss}e~7-B2, 91058 Erlangen, Germany.}
\affiliation{2 Interdisciplinary Center for Molecular Materials (ICMM) and
  Computer-Chemistry-Center (CCC), Friedrich-Alexander-Universit\"at
  Erlangen-N\"urnberg (FAU), N\"agelsbachstra{\ss}e~25, 91052 Erlangen,
  Germany.}
\affiliation{3 Intitute of Nanotechnology, Karlsruhe Institute of Technology (KIT),
  Hermann-von-Helmholtz-Platz~1, 76344 Eggenstein-Leopoldshafen, Germany.}
\affiliation{4 Max-Born Institute for Nonlinear Optics and Short Pulse
  Spectroscopy, Max-Born Stra{\ss}e 2A, 12489 Berlin, Germany.}

\date{\today}


\begin{abstract}
Parallel (``nested'') regions of a Fermi surface (FS) drive instabilities of the electron fluid, for example the spin density wave in elemental chromium. In one-dimensional materials, the FS is trivially fully nested (a single nesting vector connects two ``Fermi dots''), while in higher dimensions only a fraction of the FS consists of parallel sheets. We demonstrate that the tiny angle regime of twist bilayer graphene (TBLG) possess a phase, accessible by interlayer bias, in which the FS consists entirely of nestable ``Fermi lines'': the first example of a completely nested FS in a 2d material. This nested phase is found both in the ideal as well as relaxed structure of the twist bilayer. We demonstrate excellent agreement with recent STM images of topological states in this material and elucidate the connection between these and the underlying Fermiology.  We show that the geometry of the ``Fermi lines'' network is controllable by the strength of the applied interlayer bias, and thus that TBLG offers unprecedented access to the physics of FS nesting in 2d materials.
\end{abstract}

\maketitle


\section{Introduction}

Atomically thin materials often exhibit remarkable electronic states, in particular when such materials possess an emergent moir\'e lattice, a geometry found only in 2d materials.  One of the richest such systems is the graphene twist bilayer \cite{hass08,li10,tram10,Shallcross2010,Shallcross2013,Weckbecker2016} that, simply by tuning the rotation angle, connects the two fundamental paradigms of propagating Bloch electrons (found at large angles) and localized electrons (found at small angles). In this latter regime the band structure exhibits extremely flat bands with the Fermi velocity falling to zero at certain ``magic angles'' \cite{Weckbecker2016,Bistritzer2011a}, at which superconducting and Mott states have recently been observed \cite{Cao2018a,Cao2018b}.

In this work we demonstrate that the band structure of the small angle twist bilayer contains an additional phase, accessible by applied interlayer bias, characterized by a Fermi surface (FS) consisting of a network of regular Fermi lines. This represents the ultimate limit of ``Fermi surface nesting'', a FS topology known to be associated with charge density waves, Peierls deformations, and possible symmetry breaking of the Fermi liquid \cite{Dugdale2016}. While FS nesting is known in many materials, it is always both incomplete (i.e., only a small portion of the FS consists of nested sheets) and notoriously difficult to control \cite{Inosov2008,Dugdale2016}. In contrast, the nesting exhibited by the twist bilayer is both complete (100\% nested) and, as we show, can be fully controlled by tuning of the interlayer bias.

This ``nesting phase'' of the twist bilayer is found in a large regime of angle--field space and is, remarkably, found both for the ideal twist geometry as well as the structural dislocation network that it reconstructs to at tiny angles \cite{dai_twisted_2016,nam_lattice_2017,yoo_atomic_2019}. The finding of a robust moir\'e-induced 2d ``Fermi line'' analogy of the 1d ``Fermi dot'' topology, controllable via bias, both offers unprecedented access to the physics of FS nesting, as well as highlighting the remarkable electronic structures that can be created by moir\'e geometries and their structural dislocation networks in 2d materials.

\begin{figure}[h!]
    \centering
    \includegraphics[width=\textwidth]{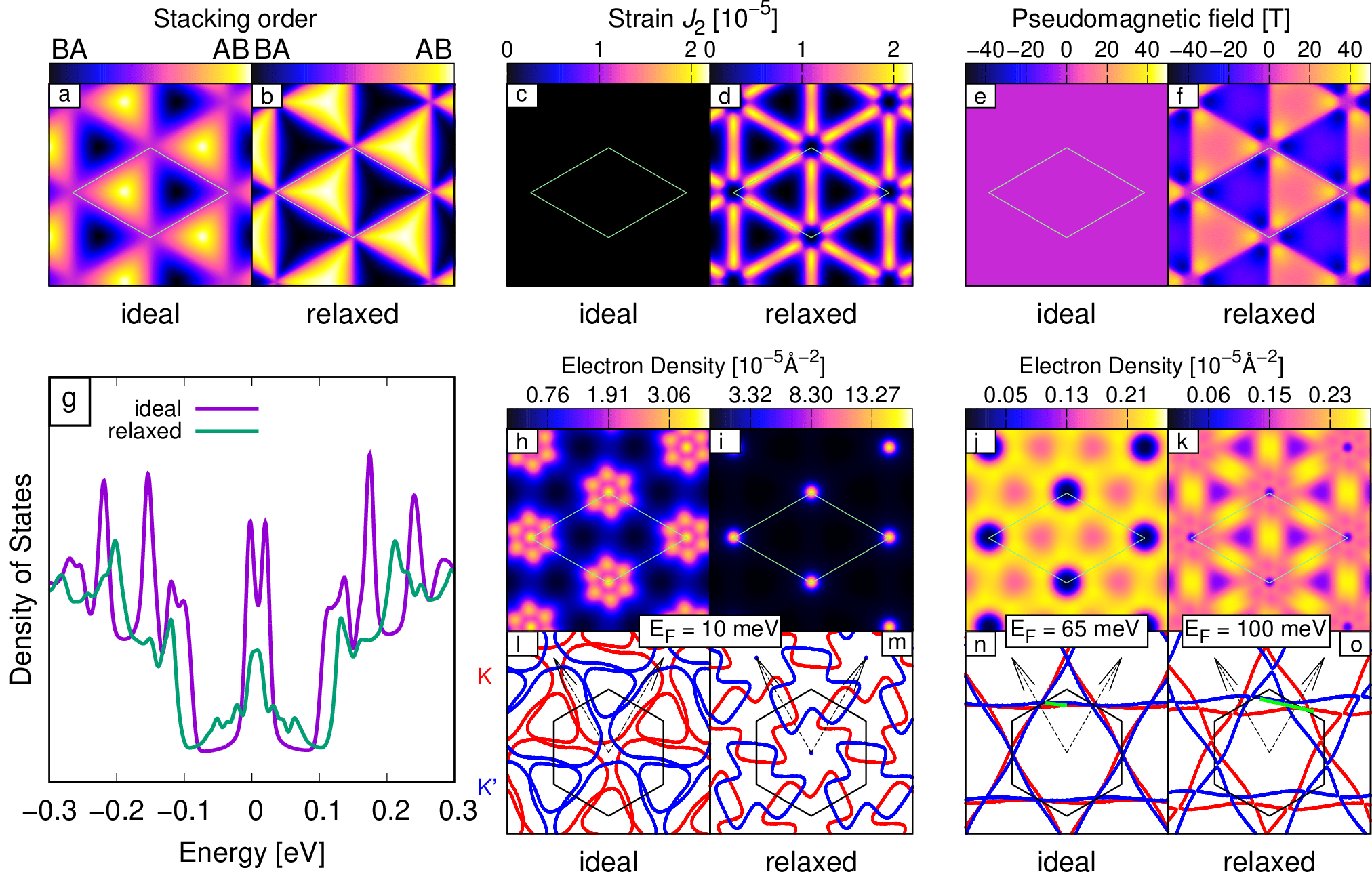}
    \caption{\emph{Fully nested Fermiology in the graphene twist bilayer and the corresponding dislocation network} ($\theta = 0.51^{\circ}$, $E=90$\,mV/$\si{\angstrom}$). Below $\sim1^\circ$ twist bilayer graphene relaxes into an ordered network of dislocations, with the smoothly varying stacking order of the ideal twist geometry (a) becoming series of sharp AB and BA domains (b), each separated by pure shear partial dislocations with high von Mises strain ($J_2$) (c,d). Pseudo-magnetic fields of the order of 40\,T are induced in the AB and BA regions with alternating sign between the latter (e,f). In the density of states (DOS) the zero mode is substantially broadened, with the valley region shifting upwards in energy (g). However, while atomic relaxation induces dramatic changes to the Fermiology in the zero mode region (l,m), in the valley region a remarkably stable Fermi topology of fully nested Fermi lines is seen (nesting vector indicated by the green line), common to both the ideal twist bilayer and dislocation network (n,o). In the local DOS atomic reconstruction strongly decreases the intensity on the AA regions close to the Dirac point (h,i), while increasing localization on the domain walls in the valley region (j,k), features matching those observed in an STM experiment of the twist bilayer (although the twist angle in the STM work, 0.245$^\circ$, is smaller than the $0.51^\circ$ studied here, we find this behaviour to be generic for dislocation networks with $\theta \lesssim 1^\circ$).
}
    \label{1}
\end{figure}

\section{Results}

\subsection{Model}

The physics of the tiny angle regime of the twist bilayer is an essentially multiscale problem involving both the lattice constant of graphene -- the scale at which atomic relaxation occurs -- and the moir\'e length of the twist unit cell, which may be many orders of magnitude greater than the lattice parameter. To capture this multiscale physics we employ a dual approach consisting of (i) an atomistic geometry optimization step mapped onto (ii) a continuum approach calculation of the electronic structure. The link between these is the relaxation field generated in step (i) that is then incorporated into the effective Hamiltonian of step (ii). This Hamiltonian consists of layer diagonal blocks, that treat single layer physics at the tight-binding level, and layer off-diagonal blocks encoding a generalized non-Abelian potential capable of describing both the twisted bilayer and the dislocation network (also sometimes refered to as structural solitons \cite{Alden2013,Butz2014,yoo_atomic_2019}) that it reconstructs to in the structure optimization \cite{Shallcross2013,Weckbecker2016}. In a bilayer geometry an out-of-plane bias field results in a $\pm V/2$ shift of the Dirac cones of each layer, with $V$ the resulting bias potential due to the electric field \cite{Zhang2009,Wang2016}. Further details of this approach are described in the Methods Section and Supporting Information.

\subsection{Fermiology at tiny angles}

\begin{figure}[t!]
    \centering
    \includegraphics[width=\textwidth]{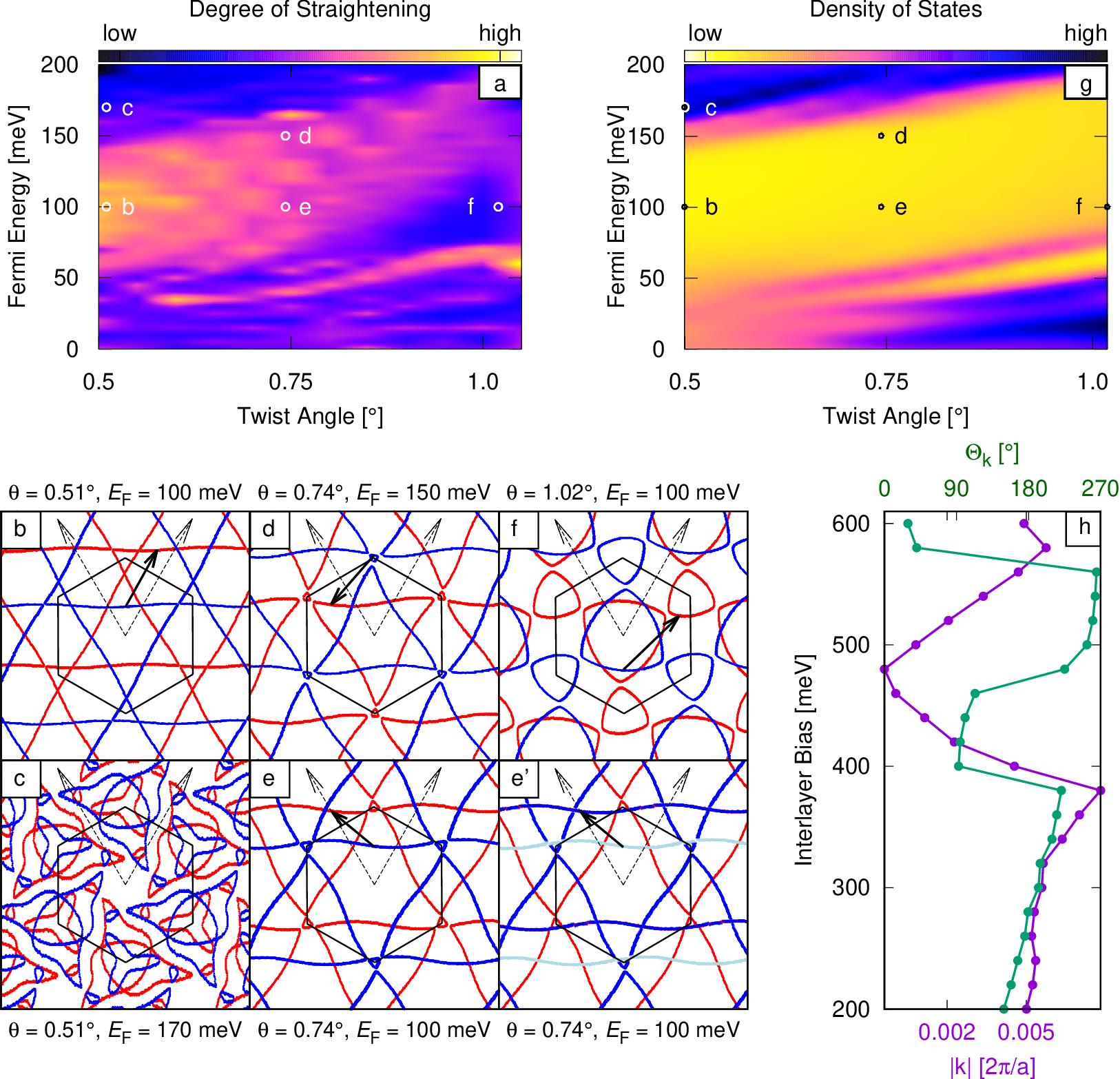}
    \caption{\emph{Nesting phase diagram of tiny angle twist bilayer graphene}: (a) Overview of the Fermi surface topology as a function of twist angle and Fermi energy at a constant electric field of $E=180$\,mV/\si{\angstrom}. Dark blue regions represent complicated non-nested Fermi surfaces (illustrated in panel c), whereas light blue and brighter regions correspond to the special fully nested Fermi network topology (panel b). The nesting region exactly matches the valley region of the density of states, see panel (b). At larger angles atomic relaxation to a dislocation network introduces a ``waviness'' to the Fermi lines (d,e) and a hybridization at the nodes of the Fermi network (f). However, the Fermi surface nesting is still perfect, as illustrated by manually shifting the horizontal blue K' Fermi line in panel (e) by the nesting vector (black arrow), which then perfectly coincides with the red K Fermi line, as shown in panel (e'). (h) Both the magnitude and direction of the nesting vector depend sensitively on the applied bias which is thus tunable by experiment and not an intrinsic material feature.
}
    \label{2}
\end{figure}

The ideal twist geometry of a spatially smooth change in stacking order (Fig.~1a) reconstructs, at small angles, into domains of AB and BA stacking separated by three sets of partial dislocations of pure screw character \cite{dai_twisted_2016,shall17,kiss15}, each characterized by one of three partial Burgers vectors (Fig.~1b). The von Mises strain (related to the shear deformation energy) is localized on the dislocations (Fig.~1d), and a large (up to 40 Tesla in magnitude) effective pseudo-magnetic field is induced by the relaxation, Fig.~1f. The consequences of this relaxation in the density of states are significant, with the well known ``zero mode'' broadened and the valley region reduced and shifted away from the Dirac point, see Fig.~1g. The local density of states (LDOS) at the Dirac point shows the expected charge localization on the AA regions of the lattice, reduced in the relaxed bilayer as these regions shrink due to their high stacking energy. In the valley region, however, a strikingly different LDOS is observed with localization now on the dislocation lines between AB/BA domains, and charge expelled from the AA regions. Similar features are seen in the ideal bilayer although with markedly reduced contrast between the AB/BA boundary regions. This localization on dislocation lines, along with the AA charge expulsion, have recently been reported in a scanning tunneling microscopy experiment \cite{Huang2018}.

Having established feature level agreement with experiment, we can now probe deeper into the underlying electronic structure of the valley region via examination of the corresponding Fermi surfaces (FS). Close to the Dirac point the ideal twist bilayer presents a very complex Fermi surface, characteristic of the strongly localized small angle limit (Fig.~1l), which upon lattice relaxation is dramatically altered (Fig.~1m). However, in the valley region a very different Fermiology is found that, remarkably, represents the ultimate limit of FS nesting: the three high velocity ``Fermi lines'' from the K valley can be completely translated into the three Fermi lines from the K' valley (Fig.~1o). In dramatic contrast to the Dirac point FS, this Fermiology is robust against lattice relaxation, demonstrating that this nesting FS network is an intrinsic feature of the small angle twist bilayer, whether in ideal geometry or relaxed. While FS nesting is well established in 3d materials, for example bcc Chromium, and in 2d materials such as TaSe$_2$, it is always incomplete, i.e. only restricted regions of the FS can be connected by a nesting vector. The $\theta=0.51^\circ$ twist bilayer, on the other hand, requires only 3 translation vectors to achieve 100\% nesting. Most strikingly, while the nesting vector is usually regarded as a fixed material property, its dependence on the presence of interlayer bias points to the possibility of controllable and complete FS nesting.

\subsection{Controllable nesting}

\begin{figure}[t!]
    \centering
    \includegraphics[width=\textwidth]{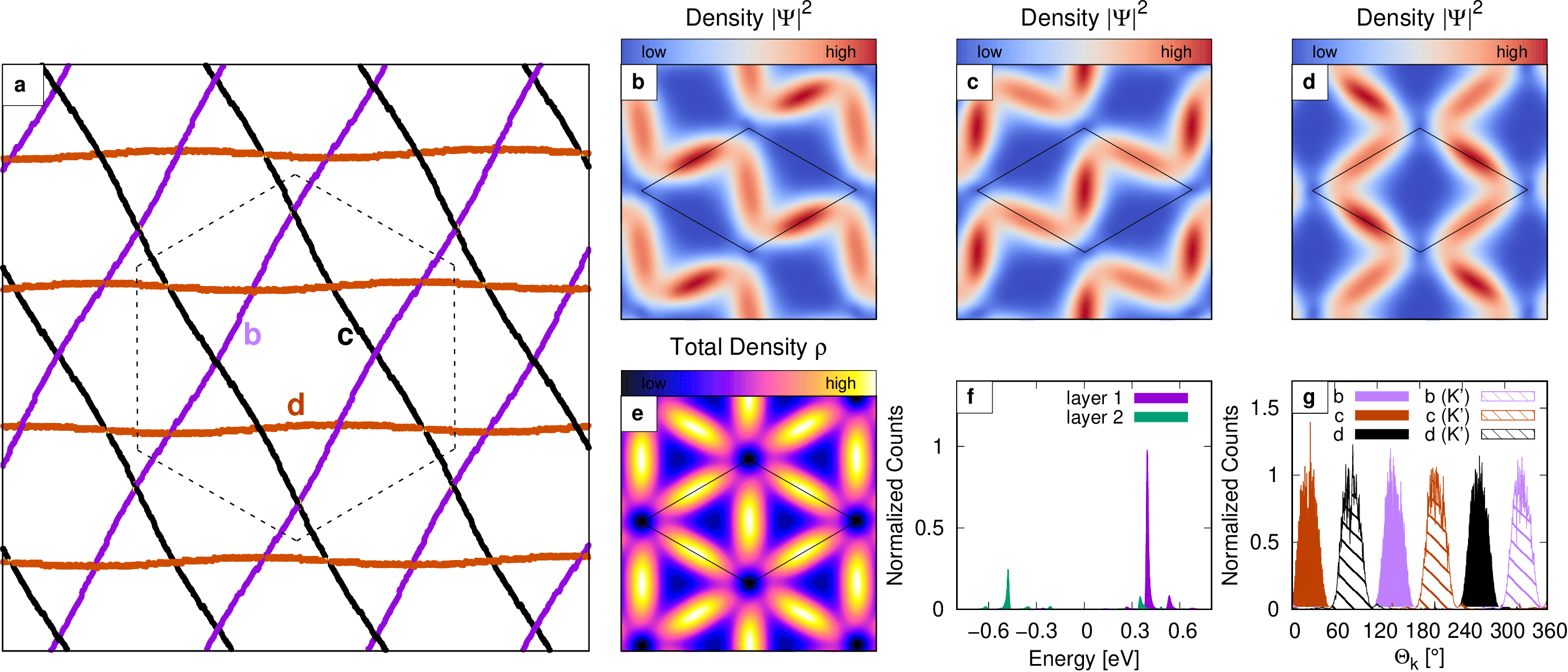}
    \caption{\emph{Electronic structure of the nesting wavefunction} ($\theta = 0.51^{\circ}$, $V=\pm0.3$\,eV, $E_F = 50$\,meV): (a) Each line of the Fermi network panel is associated with states localized predominately on one of the three pure shear partial dislocations in real space, see panels (b,c,d). Together these result in the localization on the dislocation network seen in experiment (e). Despite this complex modulation, the states in each Fermi line are remarkably close to those of single layer graphene (SLG), as shown by the SLG spectral projections (f,g). Averaged over the Fermi line, the contributing SLG states are found to have almost the same energy (f), with a momenta distribution localized in a narrow angle range which corresponds to the orientation of the Fermi line in reciprocal space (g).
}
    \label{3}
\end{figure}

A condition for such controllable nesting is that the FS topology of Fig.~1o represents the generic behavior of the twist bilayer and not an isolated special case. To examine this we create a $(\Theta,E_F)$ phase diagram of FS topology, for which a single variable characterization of the degree of straightness of the FS is required. To achieve this we create a distribution of velocity directions on the Fermi surface $N(\Theta_k)$, and measure straightness by the variance $\left<(N(\Theta_k) - \left<N(\Theta_k)\right>)^2\right>$: for a perfectly circular Fermi surface this would be zero (all directions between 0 and $2\pi$ are equally likely), while in the limiting case of a single perfectly straight Fermi line this will take its maximum value. In order to expedite the structural optimization step, for calculating the phase diagram we employ a simple structural model that well reproduces the relaxation field, see Supplementary Information.

As shown in Fig.~2a, a network of nested FS lines indeed represents the generic behaviour for the dislocation network of the reconstructed twist bilayer (it is also generic for the ideal geometry, see Supplementary Information). To illustrate the link between the phase diagram ``degree of straightening'' and FS topology we show in Figs.~2b and 2c a FS corresponding to the bright regions, which is almost perfectly straight, and one corresponding to the dark regions, which exhibits a baroque and non-nested topology. While FS nesting is robust against relaxation it both opens small gaps at the intersections of FS lines, and introduces some waviness, features more pronounced at larger angles and lower fields (Figs.~2d-f). Remarkably, these effects fully preserve the nesting feature as can be seen by the displacement of the blue K' line onto the red K line, compare panels 2e and 2e'.

Interestingly, the nesting region of the phase diagram correlates almost perfectly with the valley region in the density of states (Fig.~2g) that lies between the correlated peak at the Dirac point and the high energy shoulder regions. In this region the four bias displaced Dirac cones are altered by the interlayer interaction to 6 intersecting distorted planes, 3 from each high symmetry K point. A constant energy slice through these planes then results in the intersecting Fermi lines seen in Figs.~2b,d,e,f. The nesting vector is therefore controllable as a function of Fermi energy and, more importantly, electric field, which by altering the Dirac cone displacement alters the nesting vector. As shown in Fig.~2h the bias can tune the nesting vector through a wide range of both magnitude and directions.

\subsection{The nesting wavefunction}

As a probe of the underlying physics we now examine the nature of the nesting wavefunction; we employ the twist bilayer with $\theta=0.51^\circ$, interlayer bias $V=\pm0.3$\,eV, and Fermi energy 50~meV for which the FS is composed of perfectly nested Fermi lines (Fig.~3a). The density obtained by integrating over states from each of the three individual Fermi lines is shown in Fig.~3b,c,d, which, taken together, reproduce the localization on the dislocation network seen in experiment (Fig.~3e). However, this gross feature masks a remarkable connection between two real and reciprocal space networks: each nesting line in the FS is associated with dominant localization on one of the partial dislocations, and sub-dominant localization on a second. Furthermore, despite the seeming complexity of a localized wavefunction, the spectral decomposition of each Fermi line onto single layer graphene (SLG) states reveals an underlying simplicity: each Fermi line is dominated by states from a single energy (Fig.~3f), with the direction of the SLG states $\Theta_\bk$ centered at a single angle that corresponds to the direction of the Fermi line, Fig.~3g, with a phase shift of $\pi/3$ on going from K to K'. Unexpectedly, therefore, each Fermi line is well described by a single broadened SLG state whose velocity matches the direction in reciprocal space of the Fermi line.


\begin{figure}[t!]
    \centering
    \includegraphics[width=0.63\textwidth]{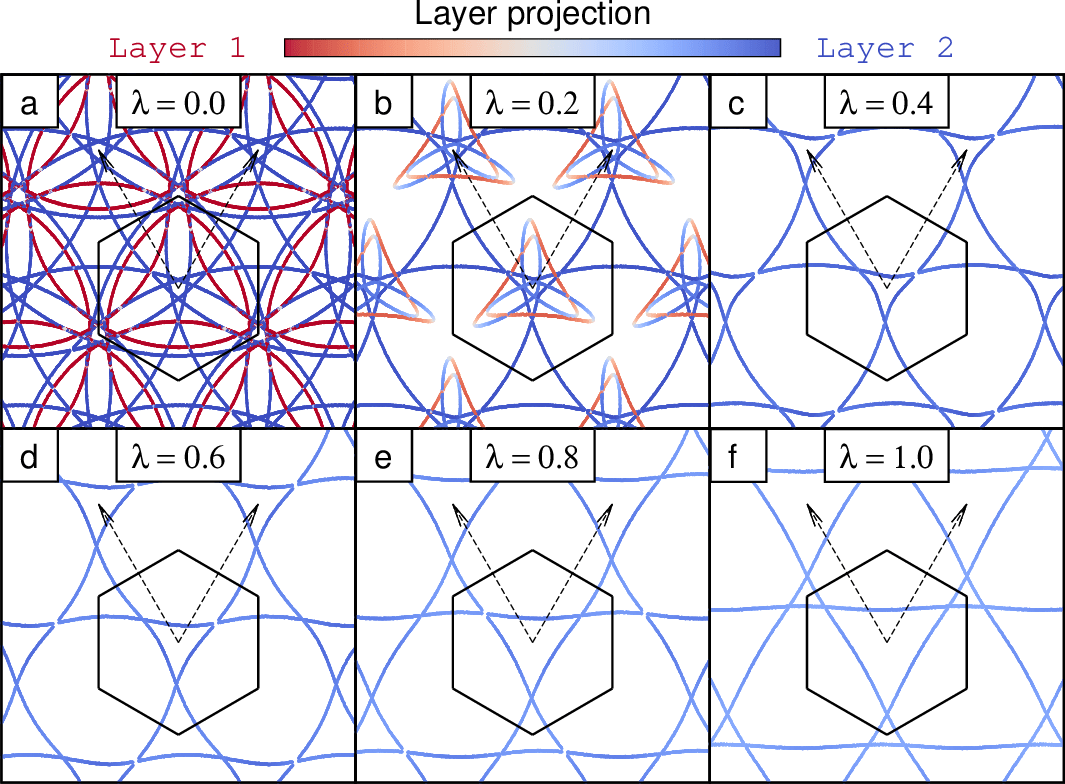}
    \caption{\emph{Emergence of the nested phase upon switching on of the interlayer interaction} $\theta = 1.02^{\circ}$, $V=0.6$\,eV, $E_F =75$\,meV: (a) For non-interacting layers the Fermi surface consists of a set of back-folded Dirac cone arcs from the individual layers. (b,c) With increasing layer coupling strength, conical arcs from each layer that are nearly coincident in momenta strongly hybridize and are removed from the energy window, leaving only the non-coincident portions of the Dirac cone (c). These are increasingly ``straightened out'' to give rise to the nested regime when the layer interaction reaches its full strength (panel f). Note that while this schematic shows only the K sheet for better visualization, the effect is identical for the K' sheet.
}
    \label{4}
\end{figure}

The accepted paradigm of the small angle twist bilayer is of massive back-folding generating multiple degeneracies at which interlayer interaction induced hybridization opens multiple mini-gaps driving the creation of very flat bands, as exemplified by the ``magic angles'' at which the Fermi velocity falls to zero at the Dirac point. This ``strong coupling'' produces Fermi surfaces that are extraordinary complex \cite{Weckbecker2016} and wavefunctions consisting of the coupling together of very many single layer states \cite{Shallcross2013}, as well as driving a physics of strong correlation. The simplicity of the nested FS and their wavefunctions found at finite bias, however, suggests a quite different physics in play for the valley region. To explore this in Fig.~4 we exhibit the creation of a FS in the valley region of the DOS as a series of ``snap shots'' as the interlayer interaction, scaled by $0 < \lambda < 1$, is switched on. With no interaction the FS is extremely complex, consisting of multiple back folded Dirac arcs, some of which are in near degeneracy (Fig.~4a). As $\lambda$ is increased these near degenerate Dirac arcs hybridize and consequently are repelled out of the valley region (Fig.~4b), both strongly enhancing the valley in the DOS as well as leaving weakly hybridized Dirac arcs, that constitute only a narrow angle segment of the original Dirac cone and have a Fermi velocity close to that of SLG. By further increase in $\lambda$ these merely ``straighten out'' to leave the nesting Fermi surface (Fig.~4c-4f). These remaining Dirac arcs are the ``helical modes'' of the AB/BA interface, recently observed in experiment \cite{Huang2018}, but whose rich Fermiology has not previously been noticed. Contradicting the conventional paradigm, the small angle twist bilayer thus contains both strong and weak coupling regimes that nevertheless have a common origin in hybridization pressure. However, in dramatic contrast to the weak coupling at large twist angles, which simply preserves the Dirac cones of each layer, small angle weak coupling results in the complete and controllable nesting of the Fermi surface.

\section{Discussion}

We have demonstrated that twist bilayer graphene, in the small angle regime and at finite interlayer bias, contains a phase characterized by Fermi surfaces of intersecting and perfectly nested Fermi lines. This Fermi surface is intrinstic to both the ideal geometry of the twist bilayer and the dislocation network it reconstructs to at small angles, a robustness in dramatic contrast to electronic structure close to the Dirac point. In contrast to all other 2d and 3d materials in which nesting is restricted to special regions of the FS, for example ``dog bone'' structures \cite{bor08,Inosov2008,Chen2016}, in small angle twisted bilayer graphene the entire FS is nested. This material therefore offers unprecedented access to the physics of FS nesting, opening many possibilities to address questions from the nature of charge density waves in 2d, to the connection between FS nesting and symmetry breaking of the Fermi liquid. Both the ``magic angles'' \cite{Bistritzer2011a} seen at the Dirac point and the complete and controllable FS nesting identified here, arise from the same physical mechanism of hybridization pressure induced by the twist back-folding, highlighting the importance of the moir\'e structure in 2d materials.

\section{Methods}

\subsection{Structure optimization}


The relaxation field $\bu_\alpha(\br)$ associated with the dislocation network in reconstructed twisted graphene bilayers was determined by atomistic force-field geometry relaxations using the LAMMPS software package \cite{LAMMPS}. The C--C interactions within the graphene layers were described by the general Amber force-field (GAFF) \cite{GAFF}. For the non-bonded interactions between the layers we used our own implementation \cite{Butz2014} of the registry-dependent Kolmogorov-Crespi potential \cite{KC2005}, which gives an accurate representation of the 3d generalized stacking-fault energy in bilayer graphene (see Supplementary Section~1 for more details).

\subsection{Electronic Hamiltonian}

The starting point for the electronic analysis is a two-center tight-binding Hamiltonian

\begin{equation}
H = \sum_{ij} t_{ij}c_i^{\dagger}c_j + \sum_i U_i c_i^{\dagger}c_i
\label{Htb}
\end{equation}
with interlayer bias added via an onsite energy shift $U_i$, with $U_i = +U$ for $i\in$~layer one and $U_i = -U$ for $i\in$~layer two. We adopt a Gaussian parameterization not restricted to nearest neighbours (nn), $t(\bdel) = A e^{-B\bdel^2}$ with $\bdel$ the hopping vector between sites $i$ and $j$. $B=1\,\text{eV}\si{\angstrom}^{-2}$ and for intralayer hopping $A=-21$~eV, giving a nn hopping of $-2.8$~eV, with $A = 0.4$~eV for interlayer hopping, giving a band gap of 0.76~eV for AB stacking. As shown in \cite{Rost2019} an exact map exists from a two-centre tight-binding Hamiltonian to a general continuum Hamiltonian $H(\br,\bp)$, with $\br$ and $\bp$ the position and momentum operators. Expressed conveniently in layer space this takes the form

\begin{equation}
H(\br,\bp) =
\begin{pmatrix}
 H^{(1)}(\br,\bp)  & S(\br,\bp) \\
S^\dagger(\br,\bp) & H^{(2)}(\br,\bp)
\end{pmatrix}
\label{Heff}
\end{equation}
where the $2\times2$ layer diagonal blocks $H^{(n)}$ are written in sub-lattice space and consist of the single layer tight-binding Hamiltonian $H_{SLG}^{(n)}(\bp)$ augmented at lowest order by scalar and pseudo-gauge fields encoding the relaxation through the strain tensor $\varepsilon_{ij} = \frac{1}{2} \left(\partial_j u_i + \partial_i u_j \right)$

\begin{equation}
 H^{(n)}(\br,\bp) = H^{(n)}_{SLG}(\bp) + \alpha_1 \sigma_0 (\varepsilon_{11}^{(n)} + \varepsilon_{22}^{(n)}) + \alpha_2 \bsig.(\varepsilon_{11}^{(n)}-\varepsilon_{22}^{(n)}, 2\varepsilon_{12}^{(n)}) + \ldots
\end{equation}
with $\bsig = (\sigma_x,\sigma_y)$ the vector of Pauli matrices and $\sigma_0$ the identity matrix. The mapping process generates also optical deformation terms \cite{gupta19}, which turn out to be zero for the relaxation field of the twist bilayer, as well as higher order terms in momentum and derivatives and powers of the deformation tensor. The interlayer coupling of these single layer blocks is given in the continuum representation by

\begin{equation}
\label{intLr}
 \left[S(\br,\bp)\right]_{\alpha\beta} = \frac{1}{A_{UC}} \sum_{i} M_{j\alpha\beta}
  e^{-i\bG_j.\bu^{(M)}(\br)} e^{-i\bK_j.(u^{(R)}_\alpha(\br)-u^{(R)}_\beta(\br))} \eta_{\alpha\beta}\left(\br,\bK_j+\bp\right)
\end{equation}
where the sum is unrestricted and over the single layer reciprocal vectors $\bG_j$, with $\bK_j$ these vectors measured from the high symmetry K point of layer 1. $\alpha$ and $\beta$ label sub-lattices of the $2\times2$ interlayer block and $\bnu_\alpha$ are the basis vectors of the unit cell. This expression contains both the twist through the acoustic moir\'e field $\bu^{(M)}=(R-1)\br$ with a length scale of $D=a/(2\sin\theta/2)$, and possible reconstruction to a dislocation network through the relaxation field $\bu^{(R)}_\alpha(\br)$. This latter field is generated from the atomistic data of the structural optimization step via bicubic interpolation. The amplitude of each term in the sum is given by the mixed space hopping function $\eta_{\alpha\beta}(\br,\bq)$, created by in-plane Fourier transforming with respect to $\delta_x$ and $\delta_y$ the interlayer hopping function evaluated at the relaxed hopping vector: $t(\bdel + \bu_\beta(\br+\bdel) - \bu_\alpha(\br))$. The out-of-plane part, which contains the modification of hopping amplitude due to out-of-plane deformation, remains in real space and is treated perturbatively, see Supplementary Information. Note that the in-plane part of the relaxation field, which locally changes the stacking order of the bilayer and is therefore a non-perturbative physical effect, is treated by exponentiation, i.e. non-perturbatively and in a similar manner to the moir\'e field itself.

Technically, we solve Eq.~\eqref{Heff} by expressing it in basis of eigenstates of the pristine single layer systems, $H^{(n)}_{SLG} \ket{\Psi^{(n)}_{i\bk}} = \epsilon^{(n)}_{i\bk} \ket{\Psi^{(n)}_{i\bk}}$ ($n$ is the layer index, $\bk$ the quasi-momentum,  and $i$ a band index). To ensure good convergence in the low energy sector ($|E| < 0.5$\,eV), on the order of $10^2$ basis function are needed for twist angles $\sim 1^{\circ}$, whereas for angles $\sim 0.5^{\circ}$, on the order of $10^{3}$ functions are required.

\subsection{Code availability}

The code used in these calculations is available upon request from the corresponding author.

\section{Data availability}

The data that support the findings of this study are available from the corresponding author upon request.



\section{Additional information}
\subsection{Competing financial interests}

The authors declare no competing financial interests.

\subsection{Acknowledgments}

The work was carried out in the framework of the SFB 953 ``Synthetic Carbon Allotropes'' and SPP 1459 ``Graphene'' of the Deutsche Forschungsgemeinschaft (DFG).

\newpage

\section{Supporting Information}

\subsection{Structure optimization}

For the ideal AB-stacked graphene bilayer we obtain with our calculational
setup (the GAFF force field \cite{GAFF} for the C--C interactions within the graphene
layers and the registry-dependent interlayer potential of Kolmogorov-Crespi \cite{KC2005})
an equilibrium lattice constant of $a_0=2.441$\,{\AA} and an interlayer
distance of $d_{\rm AB}=3.370$\,{\AA}. Shifting the graphene layers to
AA stacking increases the layer separation to $d_{\rm AA}=3.597$\,{\AA}
($+0.227\si{\angstrom}$ as compared to AB stacking). The AA-stacked bilayer has a higher energy of 4.4\,meV
per atom as compared to AB-stacking, corresponding to a stacking fault
energy of $\gamma_{\rm AA} = 54.9$\,mJ/m$^2$. In SP stacking order the
equilibrium distance of the graphene layers and the stacking fault
energy are $d_{\rm SP}=3.390$\,{\AA} (+0.020\,\AA) and
$\gamma_{\rm SP} = 7.1$\,mJ/m$^2$ (0.6\,meV per atom), respectively, in excellent agreement with ACFDT-RPA calculations of Srolovitz \emph{et al.} \cite{sor15}.

Using this setup full structural relaxations of twisted graphene bilayers were performed for six
twist angles $\theta$ (see Table\,\ref{T:cells}). By choosing two arbitrary
integer numbers $p$ and $q$, commensurate supercells for a twisted graphene
bilayer with a twist angle of
\begin{equation}
\sin\theta = \sqrt{3}\,\frac{2pq}{3q^2 + p^2}
\end{equation}
can be constructed, see \cite{Shallcross2010}. For $p=1$ the periodicity of the twisted graphene bilayer and the moir\'e superlattice coincide, i.e.\ each hexagonal (1$\times$1)
supercell (that contains \,$N_{\rm at} = 3q^2 + p^2$\, atoms and has a
lattice constant of \,$a_{\rm tblg} = \sqrt{3q^2 + p^2}\;a_0/2$\,) presents only three points of high symmetry stacking (AA, AB, and BA).

\begin{table*}[!b]
\centering
\def\tabcolsep{15pt}
\def\arraystretch{1.2}
\begin{tabular}{cccc}
\hline\hline
Twist angle $\theta$ & ($p$, $q$) & $N_{\rm at}$ & $a_{\rm tblg}$ [nm] \\
\hline
1.02$^\circ$ &  (1, 65) &   12676 &  13.74 \\
0.74$^\circ$ &  (1, 89) &   23764 &  18.81 \\
0.51$^\circ$ & (1, 131) &   51484 &  27.69 \\
0.33$^\circ$ & (1, 199) &  118804 &  42.07 \\
0.20$^\circ$ & (1, 331) &  328684 &  69.97 \\
0.10$^\circ$ & (1, 661) & 1310764 & 139.73 \\
\hline\hline
\end{tabular}
\caption{\label{T:cells}
  Twisted bilayer graphene (TBLG) structures considered in the structure
  optimizations. $N_{\rm at}$ is the number of atoms in the primitive
  hexagonal (1$\times$1) supercell of twisted graphene bilayer and
  $a_{\rm tblg}$ is the corresponding lattice parameter.}
\end{table*}

The key structural change in the geometry optimizations is the increase of the
areas with AB and BA stacking order and the decrease of the size of the areas
with AA-like stacking. The transition regions between the AB and BA domains
become sharper and they transform, in the tiny angle regime, into three sets of pure screw partial
dislocation lines connecting the AA spots with Burgers vectors of
\,$\BB_1 = [10\bar{1}0]a_0/3$\,, \,$\BB_2 = [01\bar{1}0]a_0/3$\, and
\,$\BB_3 = [1\bar{1}00]a_0/3$\,.

At the beginning of our calculations, relaxations were performed for single
(1$\times$1) supercells of the twisted bilayer. Depending on the initial
configuration two distinct relaxed equilibrium structures were obtained. The
appearance of two different local energy minima has already been
predicted by Dai \textit{et al.} based on their calculations using a continuum model \cite{dai_twisted_2016}, who termed these two structures the ``bending'' and ``breathing mode''. 
To rule out the existence of other local minima we performed a series of molecular dynamics simulations using
snapshots from the trajectories as initial configuration for further
geometry optimizations. This procedure revealed no further local energy minima, finding always either the bending or breathing mode configuration upon optimization of the trajectory snapshot.

\begin{figure}[!t]
\includegraphics[width=0.95\textwidth]{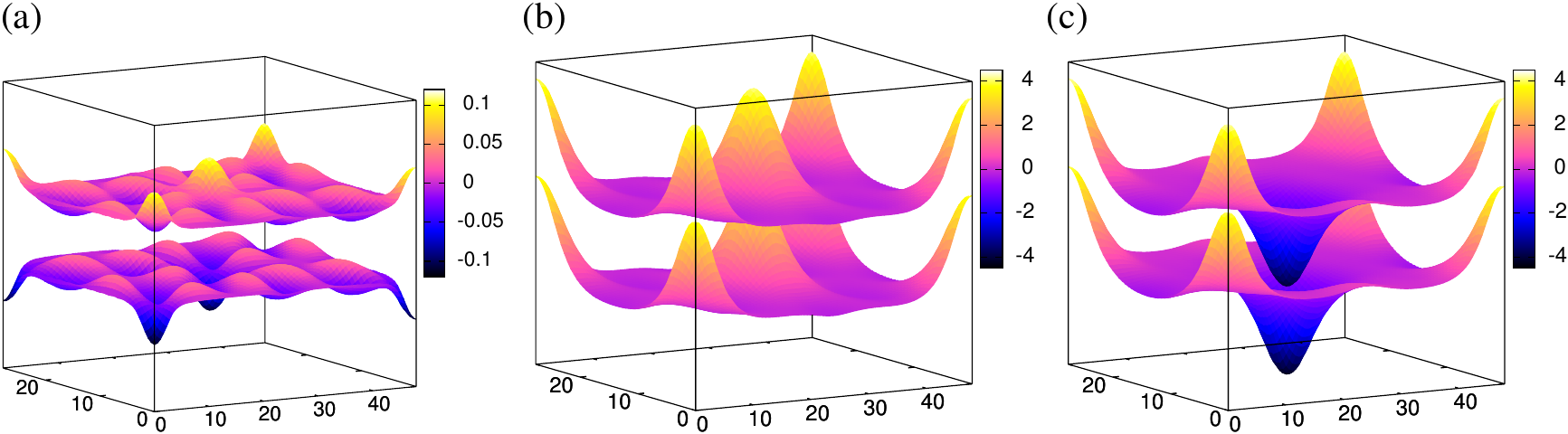}
\caption{\label{F:3d}
  Vertical (out-of-plane) relaxations of a $\theta=0.51^\circ$ twist bilayer for three local energy minima: (a) the breathing mode, (b) the bending mode with same-side AA spot bulges, and (c) with alternating-side AA spot bulges. The out-of-plane displacement $\delta$ (color-coded in {\AA}) represents the deviation from the interlayer separation of the ideal AB-stacked bilayer. The alternating-side bending mode is obtained by relaxation within an orthorhombic supercell, while structures (a) and (b) employ a hexagonal (1$\times$1) supercell. All lateral distances are in nanometers.} 
\end{figure}

These two structures differ principally in the buckling of the
graphene sheets at the AA spots. In the breathing mode the two sheets are
buckled in opposite directions (see Figure\,\ref{F:3d}a), with a very small buckling
amplitude of about $\pm$0.12\,{\AA}. This increase in
layer separation at the AA spots is essentially identical to the difference between the equilibrium interlayer spacing of an ideal AA bilayer as compared to an AB (or BA) stacked bilayer.
In the bending mode, however, both layers buckle in the same direction with an amplitude an order of magnitude larger than the breathing mode: 4.15\,{\AA} for the upper layer at a twist 
angle of $\theta=0.51^\circ$, see Figure\,\ref{F:3d}b). The physical origin of the two local energy minimum structures has been analyzed in detail in \cite{dai_twisted_2016}.

\begin{figure}[!t]
\centering
\includegraphics[width=0.55\textwidth]{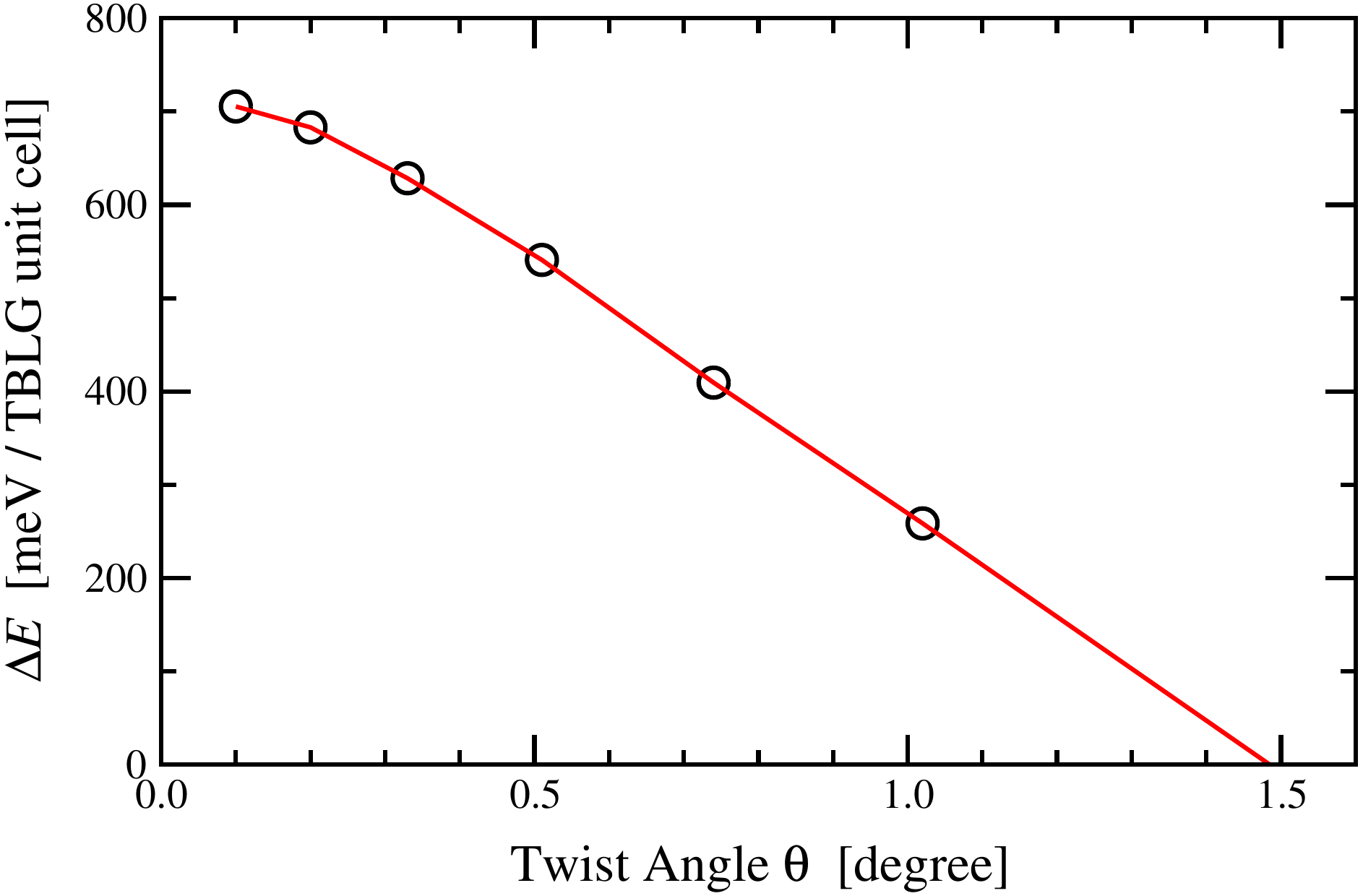}
\caption{\label{F:ebend}
  Total energy difference $\Delta E$ between the breathing and bending modes of the relaxed twist bilayer (see Figure\,\ref{F:3d}) as a function of angle. Below a turnover angle of $1.5^\circ$ the bending mode is stable, with the breathing mode stable for $\theta > 1.5^\circ$. Note however that the differences {\it per atom} are tiny with, for example, at $\theta=0.51^\circ$ a difference of only 0.01\,meV per atom between these two structures.}
\end{figure}

For small twist angles the bending mode is slightly more favorable, whereas
the breathing mode is preferred at larger angles (see Figure\,\ref{F:ebend}).
Our calculations predict the turnover point to be at about $\theta=1.5^\circ$,
closely matching the prediction of $\theta=1.6^\circ$ from the continuum model of Dai \textit{et al.} \cite{dai_twisted_2016}. However, the energy difference between the breathing and bending modes is rather small with, for example, at $\theta=0.51^\circ$ a difference of 541\,meV, or only 0.01\,meV per atom.

\begin{figure}[!b]
\centering
\includegraphics[width=0.55\textwidth]{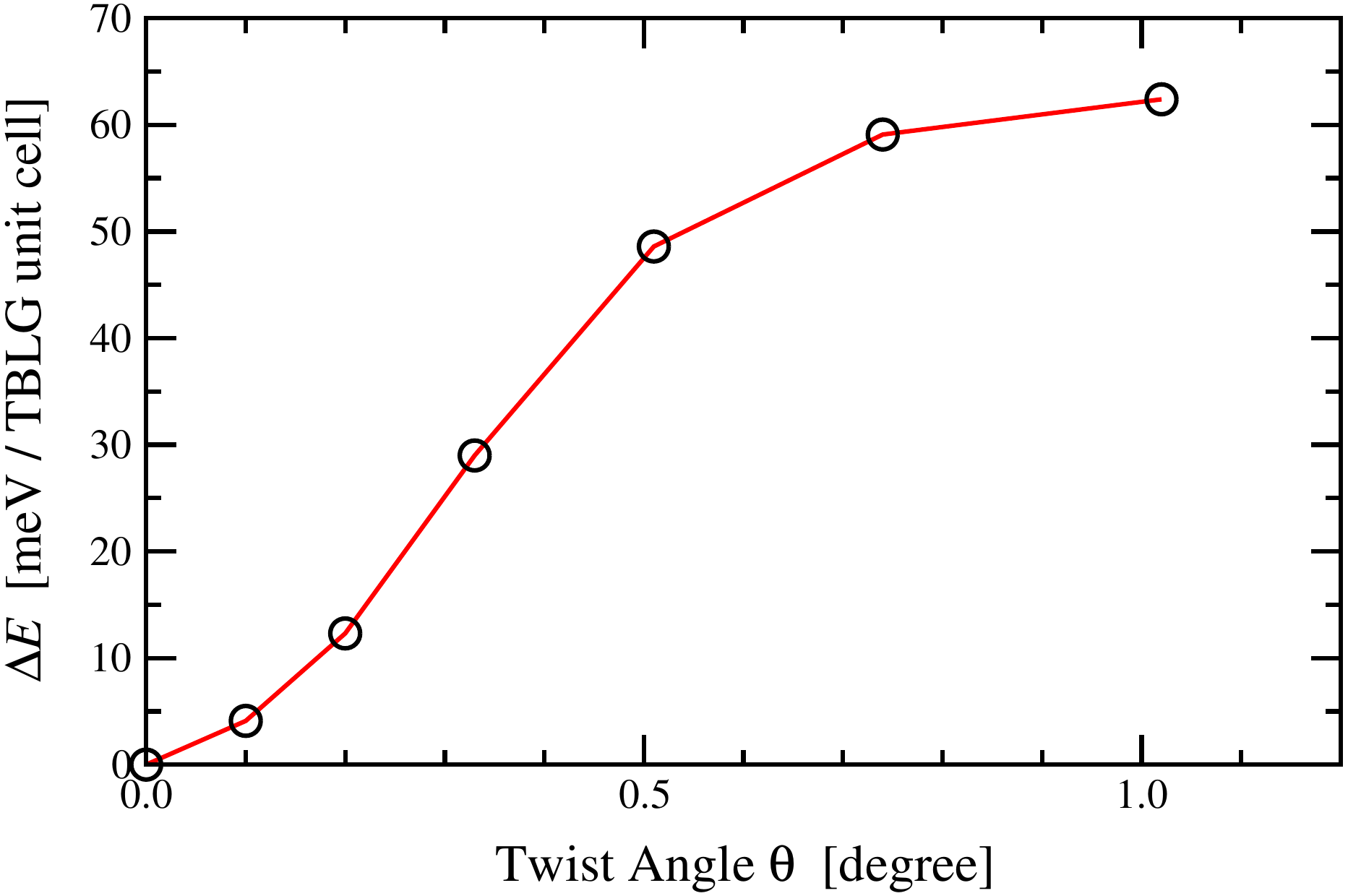}
\caption{\label{F:ortho}
  Energy difference $\Delta E$ between bending mode structures with
  same-side and alternating-side bulges normalized to one hexagonal
  (1$\times$1) TBLG unit cell.}
\end{figure}

As we have employed in our geometry optimizations a primitive (1$\times$1) supercell for the twisted graphene bilayer, the bending mode bulges must necessarily point
in the same direction (see Figure\,\ref{F:3d}b). To investigate the energetics of the bulge orientation degree of freedom, we repeated our calculations using an orthorhombic cell (dimensions of $a_{\rm tblg}$ and $\sqrt{3}a_{\rm tblg}$) that contains two moir\'e units and hence allows two bending mode types: same-side and alternating-side AA bulges (see Figure\,\ref{F:3d}c). We find that the alternating-side bulge is the preferred configuration, however by only a very small energy gain which, furthermore, vanishes with twist angle $\theta$, see
Figure\,\ref{F:ortho}. Taking again the example of a twist angle of $\theta=0.51^\circ$, 
the energy gain is 48.6\,meV per hexagonal (1$\times$1) TBLG unit cell, or less
than 0.001\,meV per atom.

\begin{figure}[!t]
\centering
\includegraphics[width=0.55\textwidth]{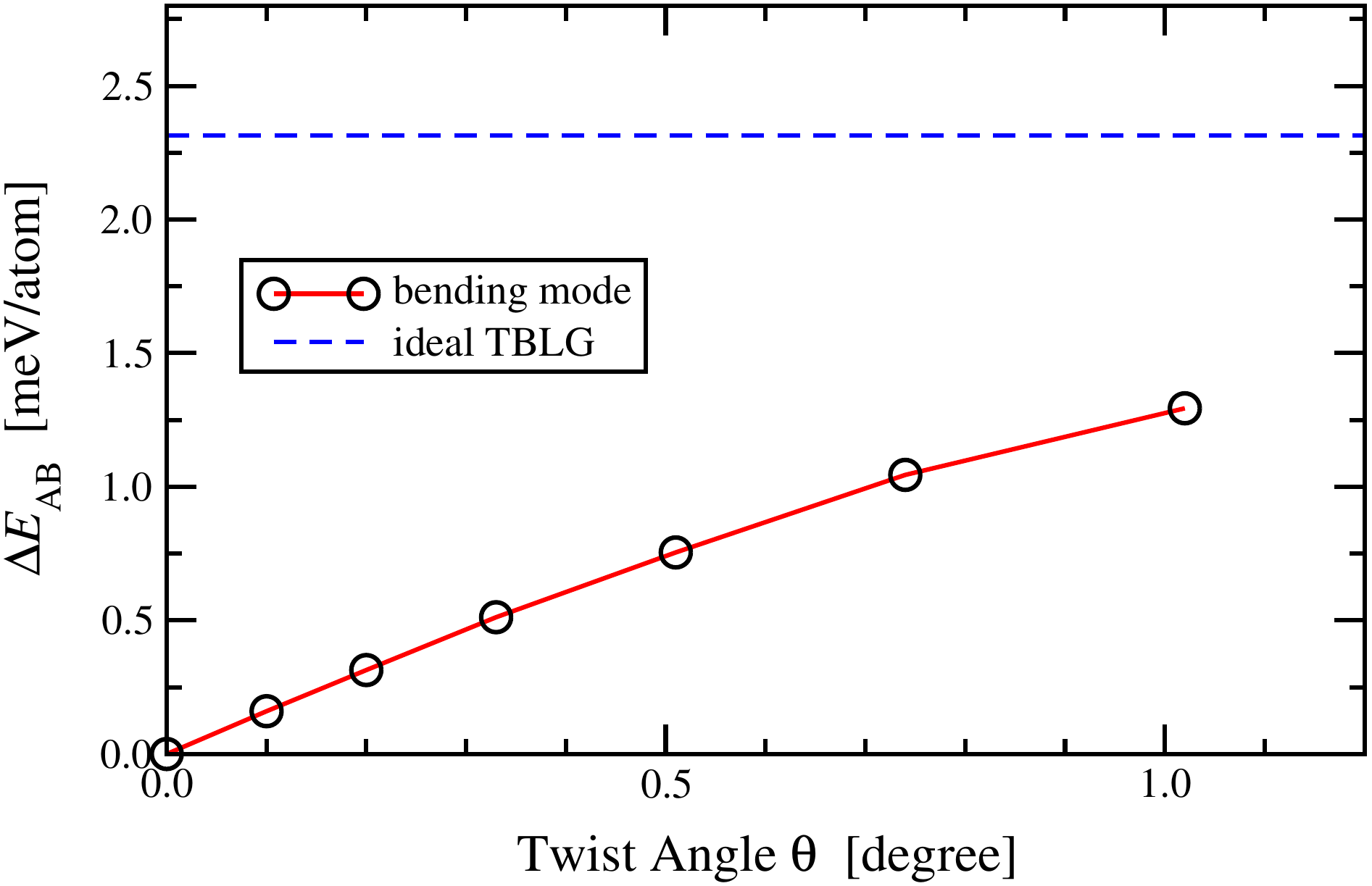}
\caption{\label{F:erel}
  Energy of the ideal moir\'e pattern and the relaxed twisted graphene bilayer
  in the bending mode structure with respect to the unrotated, AB-stacked
  bilayer (in meV per atom). On this scale the energy of the bending and
  breathing mode structures are indistinguishable. The difference between
  the ideal and the bending mode structure represents the energy gain upon
  relaxation.} 
\end{figure}

To place these energies in context we show in Figure\,\ref{F:erel} the energy difference from the ideal AB stacking of a twisted graphene bilayer with and without structure optimization. Without relaxation (i.e. in the ideal moir\'e geometry) the energy per atom is independent of twist angle, being equal to the 2d average over the generalized stacking fault energy of the graphene bilayer (2.31\,meV per atom or 28.7\,mJ/m$^2$ with our force-field setup). With relaxation, however, the energy cost for twisting a bilayer vanishes as the rotation angle approaches zero. This corresponds to a relaxation energy approaching 2.31\,meV per atom as $\theta\to 0$, of the order of 100 times greater than the difference between the breathing and bending relaxation modes.

\begin{figure}[!t]
\centering
\includegraphics[width=0.95\textwidth]{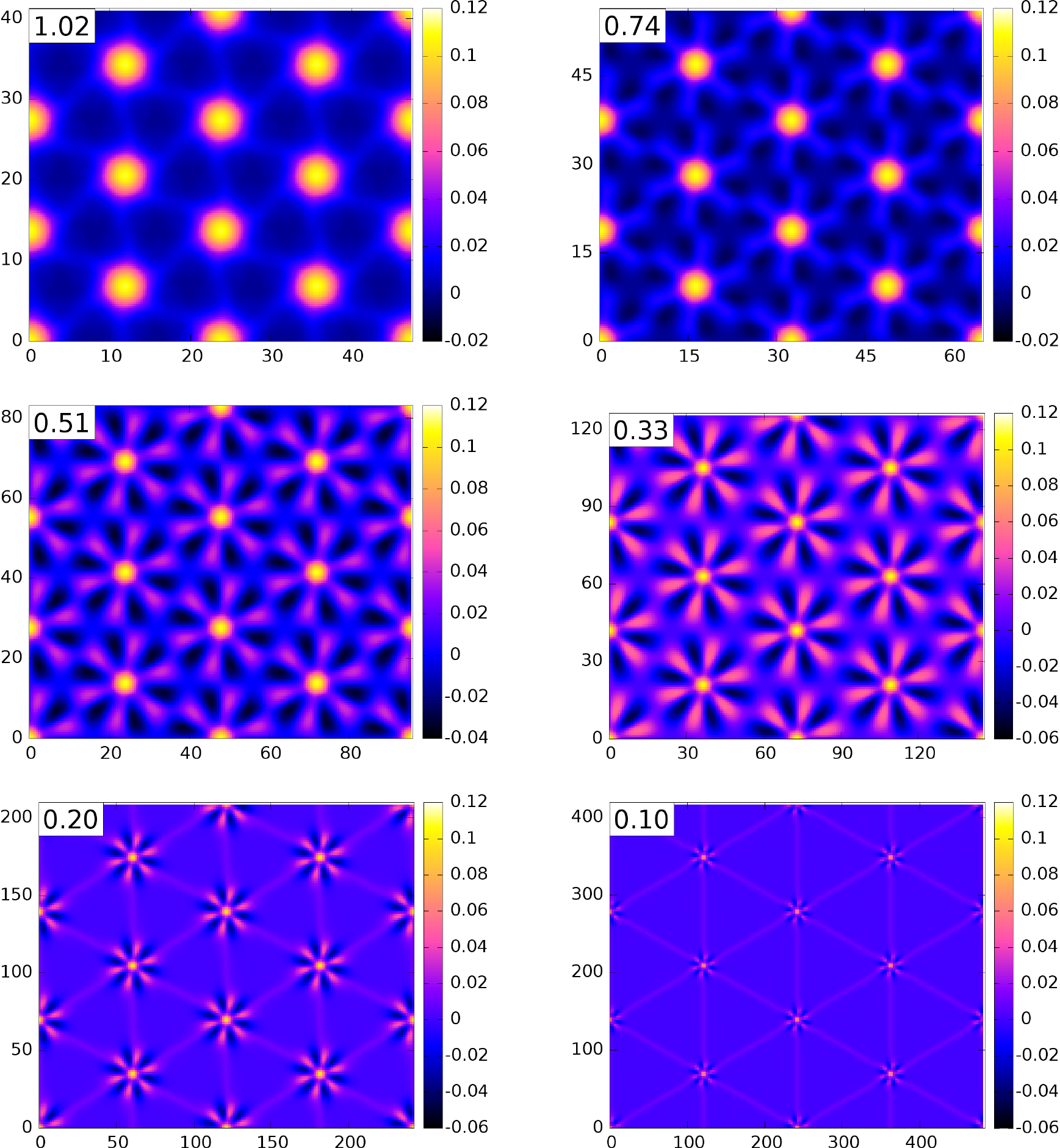}
\caption{\label{F:dz-breath}
  Vertical (out-of-plane) relaxations $\delta$ (in {\AA}) with respect
  to the unrotated AB-stacked bilayer in the breathing mode for different
  twist angles $\theta$ (in degree). All lateral distances are in nm.}
\end{figure}

\begin{figure}[!t]
\centering
\includegraphics[width=0.95\textwidth]{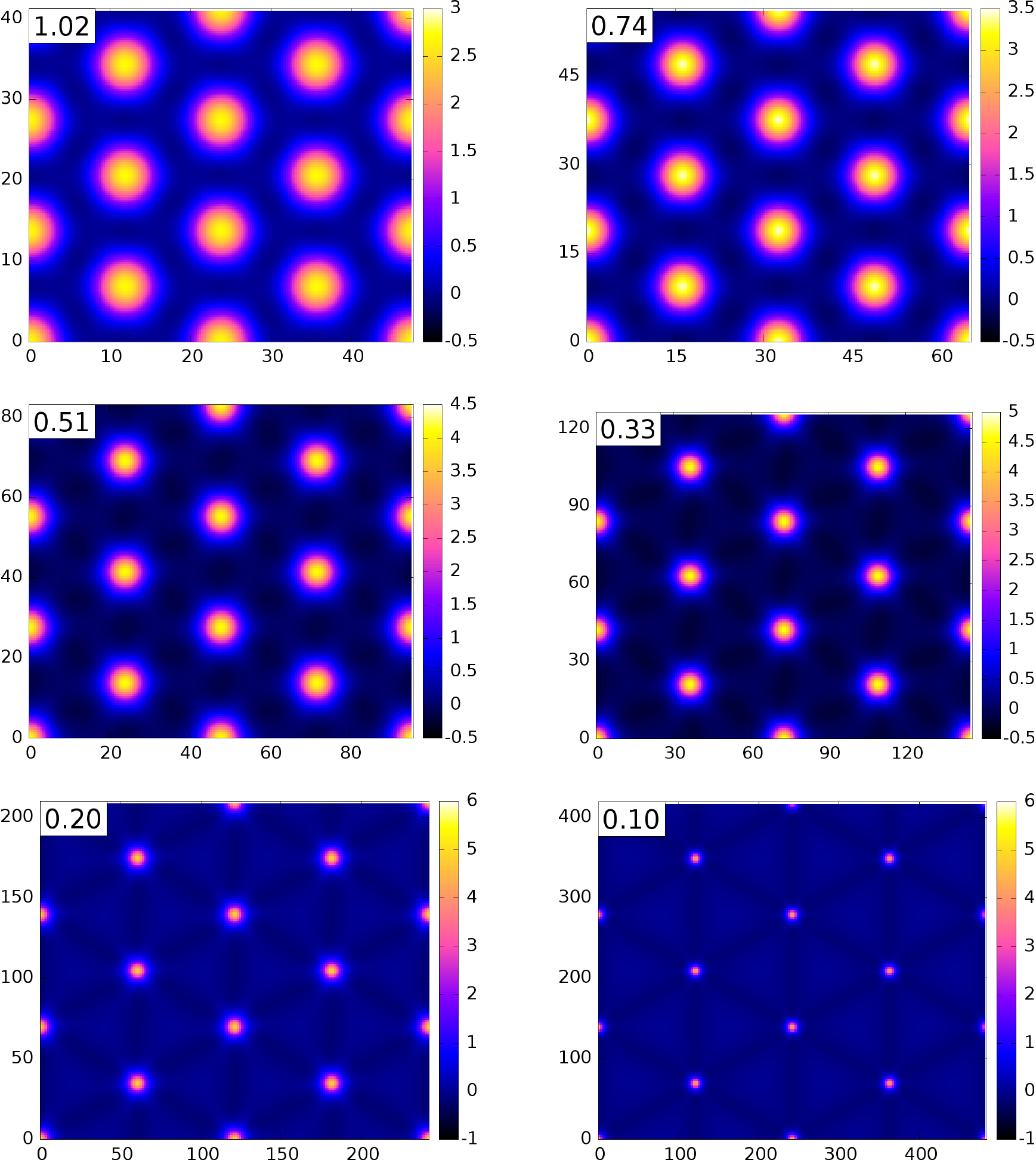}
\caption{\label{F:dz-bend}
  Vertical (out-of-plane) relaxations $\delta$ (in {\AA}) with respect to
  the unrotated AB-stacked bilayer in the bending mode for different twist
  angles $\theta$ (in degree). Only the upper layers (in the direction
  of the bulges) are shown. All lateral distances are in nm.} 
\end{figure}

Having examined the gross features of the twist bilayer relaxation as a function of angle, we now consider in detail the individual relaxed structures. In Figures\,\ref{F:dz-breath} and \ref{F:dz-bend} we show the vertical (out-of-plane) relaxation for the breathing and bending modes respectively, determined as the deviation from the ideal AB/BA stacking by subtracting the
value of $z_{\rm AB}$ in the AB/BA-stacked regions from the $z$-position
$z_\alpha$ of each atom $\alpha$ : \,$\delta_\alpha = z_\alpha - z_{\rm AB}$\,.
In the breathing mode, a peak--valley structure with 6-fold symmetry around
the AA spots emerges at small twist angles (see Figure\,\ref{F:dz-breath}), reflecting the change in stacking order as one moves away from the AA centres, and the concomitant lowering of  energy the system makes by adjusting towards the local equilibrium spacing at each local stacking order. As with all out-of-plane displacements in the breathing mode, this effect is of rather small amplitude. These structures remain confined to a radial distance of
20\,nm from the AA centres and once the separation of these centres increases beyond this distance, partial dislocation lines become visible in the out-of-plane displacement. As seen in Figure\,\ref{F:dz-breath} the out-of-plane displacement at the partial dislocations corresponds well to the difference between the ideal AB- and SP-type stacking interlayer spacing ($0.02\si{\angstrom}$). In the bending mode similar structures  are seen, Figure\,\ref{F:dz-bend}, however as the out-of-plane relaxation near the AA centres increases by more than an order of magnitude, the dislocation lines become only faintly visible as they emerge in the small angle limit.

\begin{figure}[!t]
\centering
\includegraphics[width=0.95\textwidth]{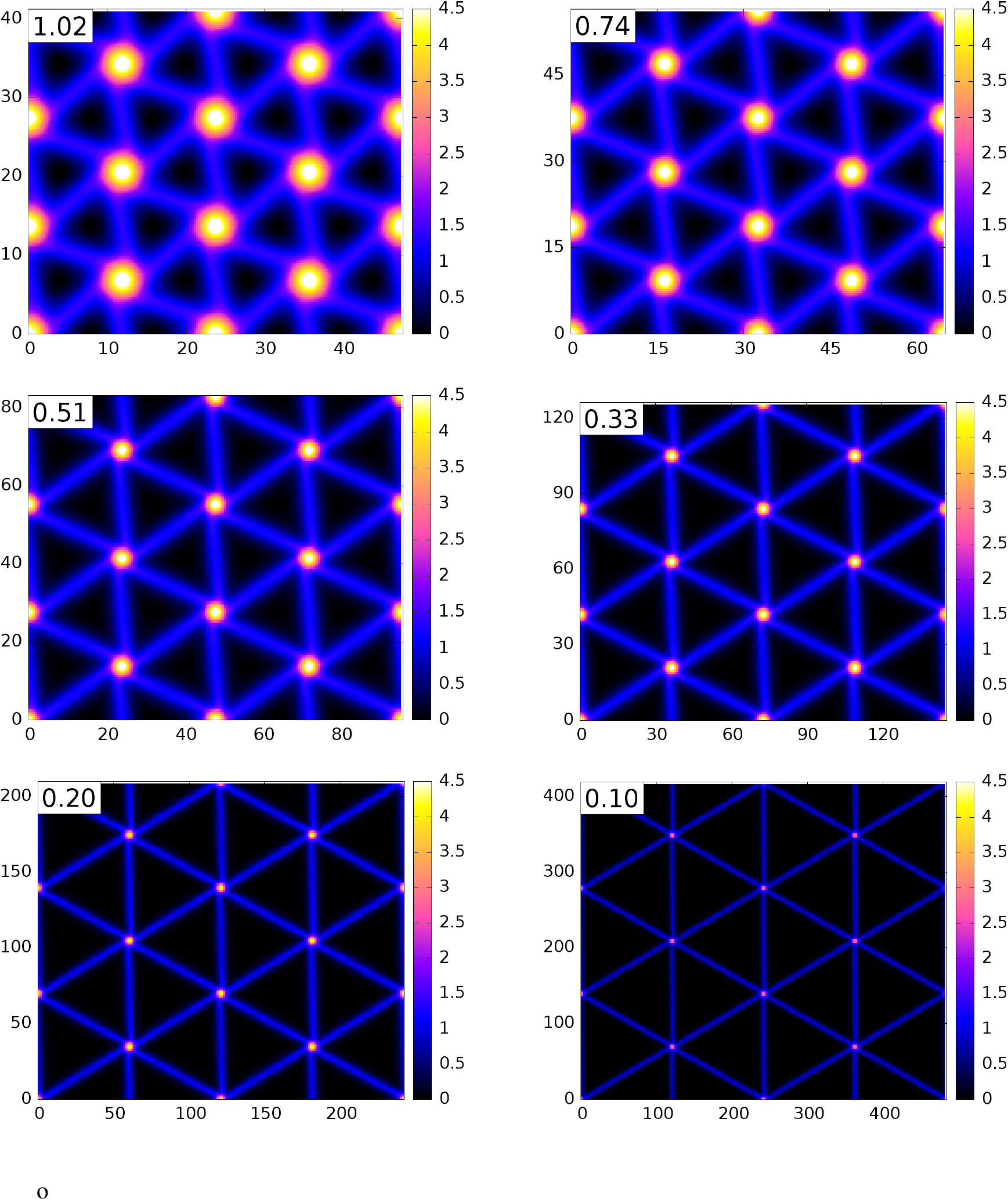}
\caption{\label{F:eat}
  Energy density distribution within twisted graphene bilayers for different
  twist angles $\theta$ (given as deviation from the unrotated AB-stacked
  bilayer in meV per atom). Only the upper layer of the bending mode is shown.
  All lateral distances are in nm.}
\end{figure}

While stable partial dislocation lines are seen only below $0.20^\circ$ in the out-of-plane relaxation a stable partial network in fact emerges at significantly larger angles (by stable we mean that a further reduction in angle changes essentially only the moir\'e length/dislocation width ratio). This may be seen from the energy per atom distribution, i.e. the deviation in the energy per atom from ideal 
AB stacking, shown in Figure\,\ref{F:eat} for the bending mode. Evidently, partial dislocation lines of 3\,nm width and AA spots of 5\,nm diameter are already stable at $1.02^\circ$. In these plots another curious relaxation feature is visible in the form of a small twist around
the AA spots, leading to a minor tilt in the orientation of the dislocation
lines. This structural feature, absent in the breathing mode, was also observed and explained by Dai \textit{et al.} in their continuum model calculations \cite{dai_twisted_2016}.

The much slower emergence of the partial dislocation network in the out-of-plane relaxation reflects the dominance of in-plane registry over out-of-plane displacement in the energy balance of the twist bilayer, and also underpins the occurrence of distinct local energy minima in the form of the breathing and bending mode. In experiment, this soft degree of freedom will likely either be pinned by environment effects, e.g. by confinement of the twist bilayer between BN layers or, if free, washed out already at very low temperatures by lattice dynamics.

\begin{figure}[!t]
\centering
\includegraphics[width=0.95\textwidth]{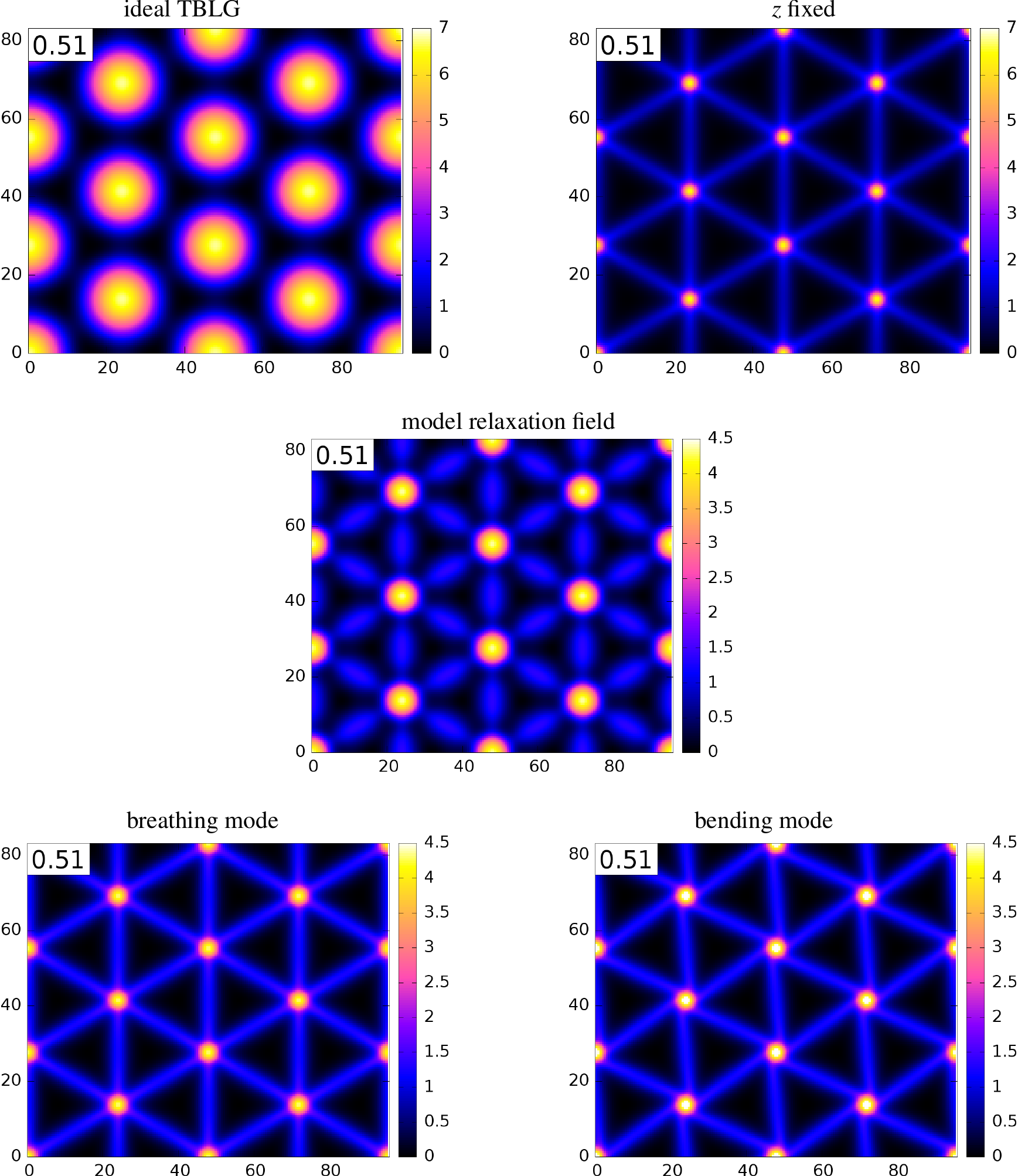}
\caption{\label{F:eat051}
  Energy density distribution for different structures of a twisted graphene
  bilayer with twist angle of $\theta=0.51^\circ$ (given as deviation from the
  unrotated AB-stacked bilayer in meV per atom). All lateral distances are in nm.}
\end{figure}

Finally, in Figure\,\ref{F:eat051} we compare the energy density distribution for the ideal twist bilayer of $\theta=0.51^\circ$ with three different relaxed structures: the bending and breathing modes, and an optimization in which the out-of-plane degree of freedom is held fixed. In all three one can observe the dramatic reduction of the AA stacking type and the emergence of the partial network. However, since the energy differences of the bending and breathing configurations is so small these distinct relaxed structures appear essentially the same (and indeed very similar to the fixed $z$ optimization), with the only feature level difference being the small tilt of the dislocation lines in the bending mode that is absent in the other two configurations. Figure\,\ref{F:eat051} also shows that the energy distribution resulting from a model relaxation field (see the next section), which evidently captures very well all relevant structural modifications and associated energy changes of the full geometry optimization.

\clearpage

\clearpage
\subsection{Model relaxation field}

\begin{figure}[hbp]
\centering
\includegraphics[width=0.95\textwidth]{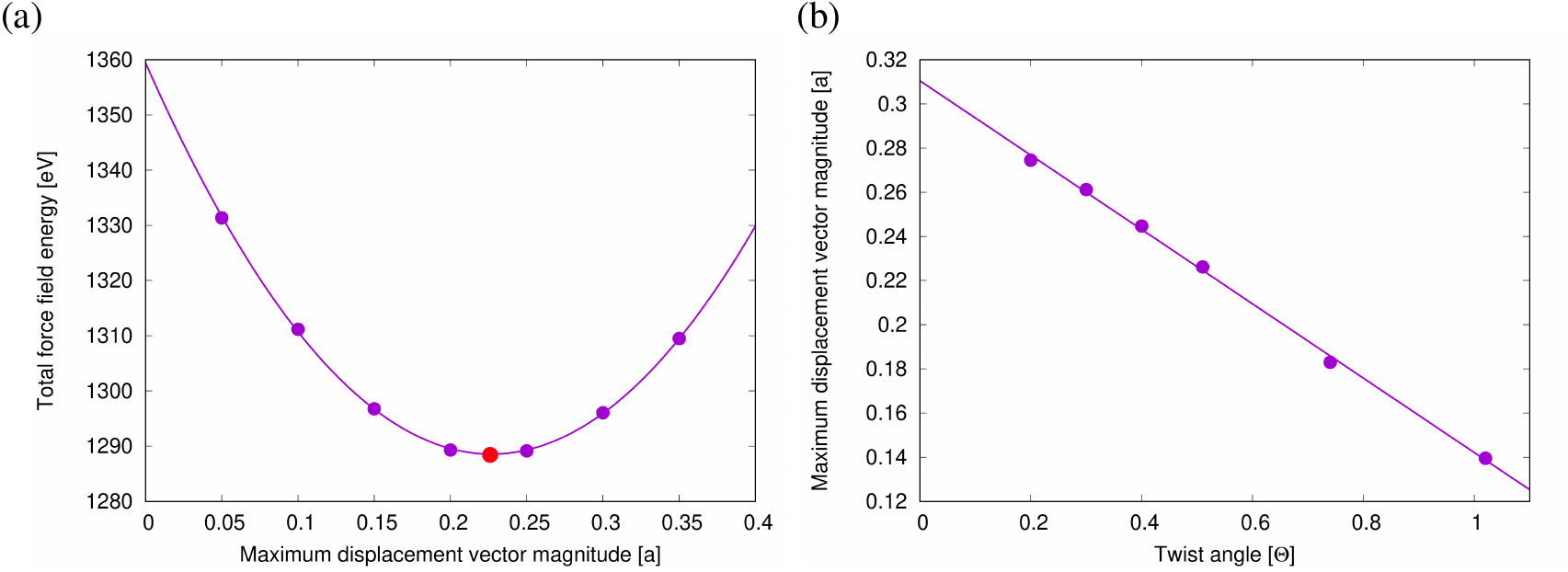}
\caption{\label{F:opt} Total energy of the twist bilayer under relaxation fields given by Eq.~\eqref{relaxeq} with different values of the relaxation field amplitude $A$. The optimum value of $A$ that minimizes the total energy is indicated by the red point. (b) The optimum value of $A$ as a function of twist angle, exhibiting a nearly linear behaviour.
}
\end{figure}

The key structural change due to lattice relaxation is the increase in the AB/BA stacking order and a corresponding reduction in other stacking types, in particular a reduction in the AA-type stacking order. At small angles $\theta \lesssim 1^\circ$ this results in a reconstruction to a triangular network of pure shear type partial dislocations separating AB/BA domains. A simple model relaxation field that captures the essence of this physics is
\begin{equation}
\mathbf{u}^l(\mathbf{r}) = \frac{1}{C_{norm}} \sum_\alpha \frac{(-1)^{l}A}{2}
(\mathbf{r}-\mathbf{R}_\alpha)^0_{\perp}|\mathbf{r}-\mathbf{R}_\alpha|
e^{-(\mathbf{r}-\mathbf{R}_\alpha)^2/B^2} 
\label{relaxeq}
\end{equation}
where $l$ indicates the layer index, the sum goes over supercell lattice points, and $(\mathbf{r}-\mathbf{R}_\alpha)^0_{\perp}$ is the vector perpendicular to the connection vector $(\mathbf{r}-\mathbf{R}_\alpha)$ in the layer plane. The sum is normalized by $1/C_{norm}$ so that the maximum amplitude of the deformation field is equal to $A$. This field generates a superposition of AA-centred displacement vector vortices of opposite orientation to the relative layer twist angle. The superposition of these deformation fields cancels at the ideal AB/BA high-symmetry positions, but away from these positions results an increase in AB/BA stacking type at the expense of other stacking types. The decay length $B$ is fixed to $1/2$ of the moir\'e length, and the parameter $A$, which determines the amplitude of the in-plane displacement vector, has been obtained by minimizing the total energy using the same force field as in the structural optimization calculations described in the previous section. A plot of $A$ versus energy can be seen in Figure\,\ref{F:opt}a (for the $\theta=0.51^{\circ}$ twist bilayer), and in this way the $A(\theta)$ can be obtained which, Figure\,\ref{F:opt}b, is to a very good approximation linear for $0.2^\circ <  \theta < 1.2^\circ$.

This simple model captures $\sim\,90$\% of the relaxation energy, as can be seen in Table\,\ref{T:Erelax}.
Some further improvement may be obtained by addition of out-of-plane relaxation, which occurs due to the layer buckling found at the AA spots in the breathing mode. This is modeled via an additional out-of-plane AA-centred Gaussian deformation whose height and width profile were again fitted to closest agreement with the structure optimization data at $1.02^{\circ}$, $0.74^{\circ}$ and $0.51^{\circ}$, and interpolated for intermediate angles. Inclusion of the out-of-plane buckling makes very little change in the electronic structure, but further improves the agreement of the relaxation energy with structure optimization as seen in Table\,\ref{T:Erelax}. 

\begin{table*}[hbp]
\centering
\def\tabcolsep{15pt}
\def\arraystretch{1.2}
\begin{tabular}{cccccc}
\hline\hline
Twist & Ideal    & In-plane     & Full         & In-plane model & Full model \\
Angle & structure        & optimization & optimization &                &            \\
 1.02$^\circ$ & 333.72   & 323.20     & 321.05   &  324.24  (90.1\%)   & 321.93   (93.0\%) \\
 0.74$^\circ$ & 625.64   & 598.32     & 595.88   &  601.26  (89.2\%)   & 598.09   (92.6\%) \\
 0.51$^\circ$ & 1355.43  & 1278.66    & 1275.66  &  1288.33 (87.4\%)   & 1283.42  (90.3\%) \\
\hline
\hline
\end{tabular}
\caption{\label{T:Erelax} Total energies of the twist bilayer for the ideal and relaxed structures, and the corresponding total energies of the model relaxation field. In the first column is shown the total energy (all energies in eV) of the ideal twist bilayer with, in the subsequent two columns, the total energy that results from structure optimization with atoms constrained to lie in the plane and a full optimization including the out-of-plane direction. The final two columns show the corresponding total energies from the model relaxation field, with the relaxation energy captured by the model shown in parenthesis.
  }
\end{table*}

\clearpage
\subsection{Theoretical treatment of lattice relaxation}

\begin{figure}[hbp]
\centering
\includegraphics[width=0.48\textwidth]{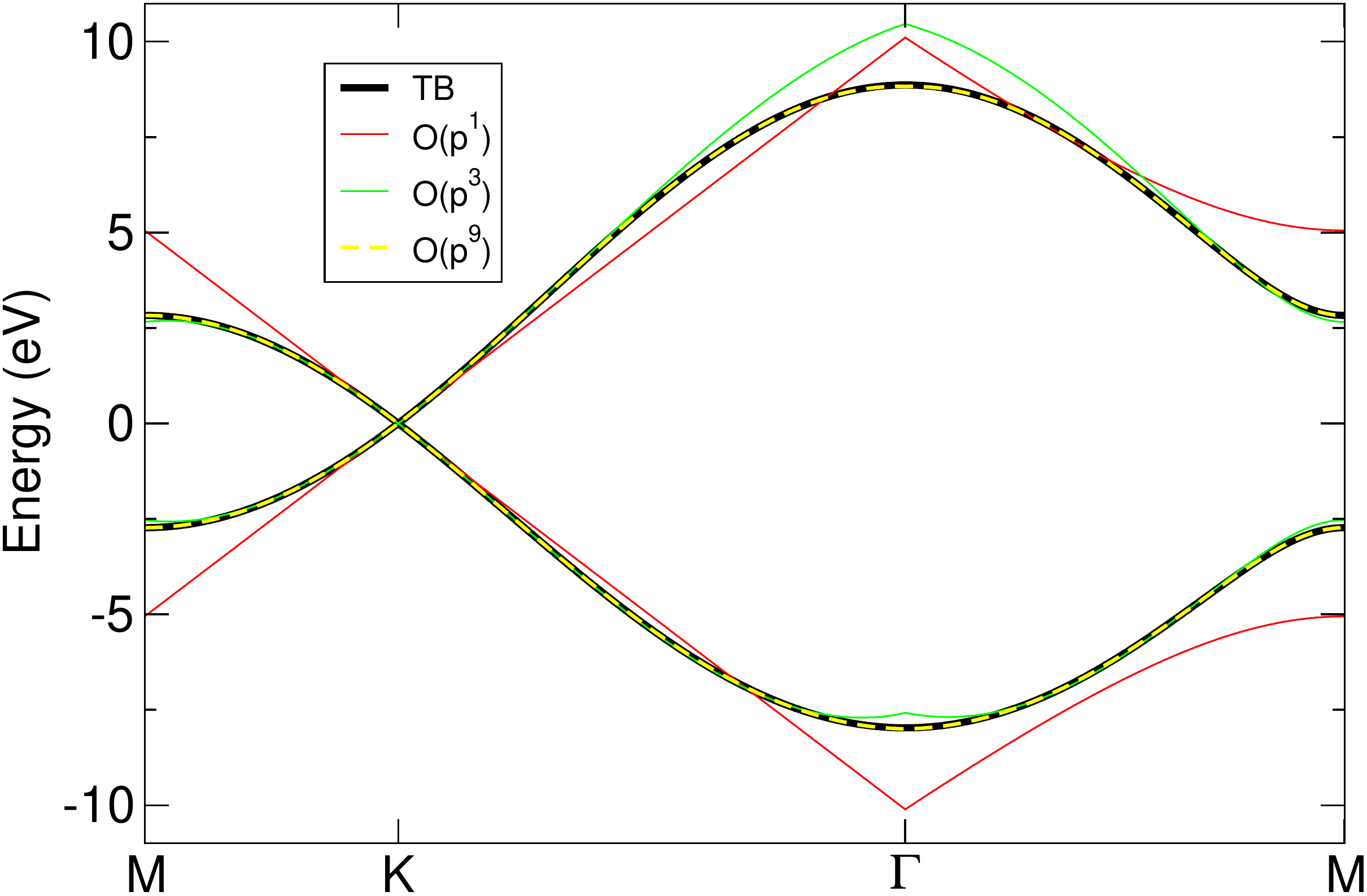}
\caption{\label{F:bnd}
Band structure of graphene calculated using the tight-binding method in the H\"uckel approximation and using the continuum approach with three different values of the momentum cutoff: $O(p^1)$ i.e. the Dirac-Weyl equation, order $O(p^3)$ which includes trigonal warping corrections, and $O(p^9)$ which shows almost perfect agreement with tight-binding over the full $\pi$-band.
}
\end{figure}

The methodology we employ here is based on the fact that an exact map exists from a general tight-binding Hamiltonian onto a general continuum Hamiltonian. Given a TB Hamiltonian

\begin{equation}
 H_{TB} = \sum_{ij} t_{ij} c^\dagger_j c_i
\end{equation}
then there exists a $H(\br,\bp)$, with $\br$ and $\bp$ the position and momentum operators, such that all matrix elements of $H_{TB}$ and $H(\br,\bp)$ satisfy the identity

\begin{equation}
\mel{\Psi_X}{H_{TB}}{\Psi_{X'}} = \mel{\phi_{X}}{H(\br,\bp)}{\phi_{X'}},
\label{opequiv0}
\end{equation}
with $X$ some quantum numbers and $\ket{\Psi_{X}}$ the Bloch functions of a suitable high-symmetry system (e.g. for a twist bilayer the AB or AA stacked bilayer), and $\ket{\phi_{X}}$ corresponding plane wave states. This approach has been fully described in \cite{Rost2019}, and used to treat optical deformations in graphene in \cite{gupta19}, and partial dislocations in bilayer graphene in \cite{kiss15,shall17,Weckbecker2019}. In what follows we describe in overview the methodology, in particular how it pertains to lattice relaxation in the twist bilayer, but refer the reader to these references for further details.

The effective Hamiltonian employed to treat twist bilayer graphene can conveniently be written in layer space has the form
\begin{equation}
H(\br,\bp) =
\begin{pmatrix}
 H^{(1)}(\br,\bp)  & S(\br,\bp) \\
S^\dagger(\br,\bp) & H^{(2)}(\br,\bp)
\end{pmatrix}
\label{HLB}
\end{equation}
The diagonal blocks describe the single layer graphene (SLG) layers that constitute the bilayer, with the coupling of these layers encoded in the $S(\br,\bp)$ blocks. The interlayer physics is driven by the local stacking order, which is both non-perturbative (the difference between the AB and AA low energy spectrum is profound) and decisively changed by reconstruction to a partial dislocation network at small angles.

The key object is $t_{\alpha\beta}(\br,\bdel)$ that describes the amplitude of electron hopping from position $\br$ on sub-lattice $\alpha$ to position $\br+\bdel$ on sub-lattice $\beta$. For the single layer blocks the exact map produces the result

\begin{equation}
 \left[H(\br,\bp)\right]_{\alpha\beta} = \frac{1}
 {A_{UC}} \sum_{j} M_{j\alpha\beta}\, \eta_{\alpha\beta}(\br,\bK_j + \bp)
 \label{Heff2}
\end{equation}
where $A_{UC}$ is the unit cell area, the sum is over reciprocal lattice vectors $\bG_j$ and $\bK_j = \bG_j - \bK_0$ with $\bK_0$ the position in the Brillouin zone from which momentum $\bp$ is measured from. The function $\eta_{\alpha\beta}(\br,\bq)$ is the Fourier transform of the hopping envelope function $t_{\alpha\beta}(\br,\bdel)$

\begin{equation}
\eta_{\alpha\beta}(\br,\bq) = \int
 \!\!d\bdel \,\, e^{i\bq.\bdel} t_{\alpha\beta}(\br,\bdel)
\end{equation}
while the ``M matrices'' are given by

\begin{equation}
 M_{j\alpha\beta} = e^{i\bG_j.(\bnu_\alpha-\bnu_\beta)}
 \label{Mmat}
\end{equation}
with $\{\bnu_\alpha\}$ the basis vectors of the unit cell.

Application of deformation fields $\bu_\alpha(\br)$ to a high symmetry system ($\alpha$ is the basis vector index) results in change of hopping vector $\bdel \to \bdel + \bu_\beta(\br+\bdel) - \bu_\alpha(\br)$ and a concomitant change in the hopping amplitude of

\begin{equation}
 t_{\alpha\beta}(\br,\bdel) = t_{\alpha\beta}(\bdel + \bu_\beta(\br+\bdel) - \bu_\alpha(\br))
 \label{dhop}
\end{equation}
For a single layer system we may expand Eq.~\eqref{dhop} as

\begin{equation}
 t_{\alpha\beta}(\br,\bdel) = t^{(0)}(\bdel) + t^{(1)}(\bdel) \left( \varepsilon_{xx}^{(n)}(\br) \delta_x^2 + \varepsilon_{yy}^{(n)}(\br) \delta_y^2 + 2\varepsilon_{xy}^{(n)}(\br) \delta_x \delta_y \right) + \ldots
 \label{exp}
\end{equation}
with the first term the hopping function of the pristine lattice (hence no $\br$ dependence) and the second term the change in hopping due to a relaxation field resulting in a strain tensor $\varepsilon_{ij} = 1/2(\partial_i u_j + \partial_j u_i)$.

For pristine single layer graphene Figure\,\ref{F:bnd} shows the convergence to the TB band structure as the order of momentum retained a Taylor expansion of Eq.~\eqref{Heff2} with respect to $\bp$ is increased; evidently while for $O(p^1)$ (i.e. the Dirac-Weyl equation) agreement is found only at low energies, for large orders the TB is essentially exactly reproduced for the full $\pi$-band, as should be the case as the underlying map is exact. Here we use a hopping envelope function $t(\bdel) = A e^{-B\bdel^2}$, with $A=-21$~eV and $B=1\,\si{\angstrom}^{-2}$ (giving a nearest neighbour hopping of 2.8~eV).

With deformation, Eq.~\eqref{Heff2} generates a series of corrections to the Hamiltonian of the ideal lattice \cite{gupta19}, the lowest order of which are the well known scalar and pseudo-gauge fields generated by the second term in the expansion shown in Eq.~\eqref{exp}. This results in the single layer Hamiltonian

\begin{equation}
 H^{(n)}(\br,\bp) = H^{(n)}_{SLG}(\bp) + \alpha_1 \sigma_0 (\varepsilon_{xx}^{(n)}(\br) + \varepsilon_{yy}^{(n)}(\br)) + \alpha_2 \bsig.(\varepsilon_{xx}^{(n)}(\br)-\varepsilon_{yy}^{(n)}(\br),  2\varepsilon_{xy}^{(n)}(\br)) + \ldots
 \label{Hd}
\end{equation}
where $H^{(n)}_{SLG}(\bp)$ exactly reproduces the TB band structure of graphene.

This latter fact suggests that a strategy for solving Eq.~\eqref{HLB} is via a basis of tight-binding eigenstates from each layer
$H^{(n)}_{SLG} \ket{\Psi^{(n)}_{i\bk}} = \epsilon^{(n)}_{i\bk} \ket{\Psi^{(n)}_{i\bk}}$ ($n$ is the layer index, $\bk$ the quasi-momentum,  and $i$ a band index), leading to Hamiltonian matrix elements of the form

\begin{equation}
 \left[H\right]_{n'i' \bk' n i \bk} = \delta_{n' i' \bk' n i \bk} \left(\epsilon^{(n)}_{i \bk} + \mel{\Psi^{(n)}_{i\bk}}{H^{(n)}(\br,\bp)}{\Psi^{(n)}_{i\bk}} \right) + \left(1-\delta_{n' i' \bk' n i \bk}\right)\mel{\Psi^{(n')}_{i'\bk'}}{S(\br,\bp)}{\Psi^{(n)}_{i\bk}}
 \label{basis}
\end{equation}
where $H^{(n)}(\br,\bp)$ contains the effect of lattice relaxation in the layer diagonal blocks, included perturbatively via Eq.~\eqref{Hd}, and $S(\br,\bp)$ contains the impact of the twist and relaxation on the interlayer interaction in the off-diagonal blocks. For these blocks the change due to deformation cannot be treated perturbatively. Fortunately, however, the Fourier transform of the hopping function can be treated \emph{exactly} at zeroth order:

\begin{equation}
 \eta_{\alpha\beta}(\br,\bq) = 
 e^{-i\bq.(\bu_\beta(\br)-\bu_\alpha(\br))} \int d\bdel\, e^{i\bq,\bdel}
 t_{\alpha\beta}(\br,\bdel)
 \label{FT}
\end{equation}
As $\alpha$ and $\beta$ are sub-lattices in different layers, $\bu_\beta(\br)-\bu_\alpha(\br)$ is the local displacement at $\br$ of the two sub-lattices from the high symmetry structure. By treating this term exactly in the Fourier transform of Eq.~\eqref{FT}, the problem of treating the change in stacking order non-perturbatively is solved.
This zeroth order term can be broken into a moir\'e field $\bu^{M}(\br) = (R-1)\br$ with $R$ the rotation operator, which describes the change from the high symmetry bilayer to the \emph{ideal} twist bilayer, and a field due to lattice relaxation $\bu^{(R)}(\br)$

\begin{equation}
 \bu_\beta(\br) - \bu_\alpha(\br) = \bu^{(M)}(\br) + \bu_\beta^{(R)}(\br) - \bu_\alpha^{(R)}(\br)
\end{equation}
The moir\'e field is purely acoustic, however the relaxation field includes both acoustic and 3 possible optical modes.
These non-perturbative parts yield phase terms in the corresponding continuum representation

\begin{equation}
\label{intLr1}
 \left[S(\br,\bp)\right]_{\alpha\beta} = \frac{1}{A_{UC}} \sum_{i} M_{j\alpha\beta}
  e^{-i\bG_j.\bu^{(M)}(\br)} e^{-i\bK_j.(u^{(R)}_\alpha(\br)-u^{(R)}_\beta(\br))} \eta_{\alpha\beta}\left(\br,\bK_j+\bp\right)
\end{equation}
while the perturbative part remains in the function $\eta_{\alpha\beta}(\br,\bq)$ and include the changes to the hopping amplitude resulting from out-of-plane lattice relaxation. In contrast to the single layer blocks, where out-of-plane deformation is a higher order correction to single layer graphene (as mirror symmetry requires it to enter as a square), the interlayer optical deformation results in a linear order correction

\begin{equation}
 \eta_{\alpha\beta}(\br,\bq) \sim
 \hat{t}^{(0)}_{\alpha\beta}(\bq) + \hat{t}_{\alpha\beta}^{(1)}(\bq) (2 d_{int}) \delta z(\br)
 + \hat{t}_{\alpha\beta}^{(1)}(\bq) \left(\varepsilon_{xx}^{(n')}(\br) q_x^2 + \varepsilon_{yy}^{(n')}(\br) q_y^2 + 2\varepsilon_{xy}^{(n')}(\br) q_x q_y\right)+\ldots
\label{2M}
\end{equation} 
with $d_{int}$ the equilibrium interlayer separation and where $n$ is the layer that sub-lattice $\beta$ belongs to and

\begin{equation}
\hat{t}^{(n)}_{\alpha\beta}(\bq) = \int \!\!d\bdel\, e^{i\bq.\bdel} \frac{\partial^n t^{(0)}_{\alpha\beta}(\bdel^2)}{\partial (\bdel^2)^n}
\end{equation}
Eqs.~\eqref{Hd} and \eqref{intLr1} together with the single layer basis, Eq.~\eqref{basis}, form a complete description of both the ideal twist bilayer, including the effect of subsequent relaxation. While retaining lattice relaxation in the interlayer interaction is obviously essential, in the single layer blocks one can envisage different levels of theoretical completeness: the pristine layer may be treated at the TB or DW level, and relaxation can be included via pseudo-magnetic and scalar fields in both these cases.

\noindent
Therefore, four different levels of theory to treat these blocks can be considered:

\begin{itemize}
 \item $H^{(n)}(\br,\bp)$ can be approximated by the Dirac-Weyl (DW) equation or
 \item $H^{(n)}(\br,\bp)$ can be treated at the tight-binding (TB) level
 \item $H^{(n)}(\br,\bp)$ can be approximated by the Dirac-Weyl equation augmented by scalar and pseudo-gauge fields describing the lattice relaxation
 \item Finally $H^{(n)}(\br,\bp)$ can be treated at the tight-binding level including these fields
\end{itemize}

As shown in Figure\,\ref{Rel}, there is very little difference in the DOS and Fermi surfaces (at low energies) between the DW and TB descriptions of the single layer blocks; there is some increased particle-hole asymmetry, but the interlayer field $S(\br,\bp)$ in any case breaks this symmetry. However, substantially more change is seen on augmenting either the DW or TB descriptions of perfect SLG by the scalar and pseudo-magnetic fields that result from lattice relaxation.

\clearpage
\subsection{Electronic consequences of relaxation}

The four distinct sub-lattices of bilayer graphene yield four relaxation modes

\begin{equation}
 \bv = \frac{1}{4} \begin{pmatrix} 1 & 1 & 1 & 1 \\ 1 & 1 & -1 & -1 \\ 1 & -1 & 1 & -1 \\ 1 & -1 & -1 & 1 \end{pmatrix} \bu
\end{equation}
with $\bu$ a four-vector of the four relaxation fields $\bu_\alpha(\br)$. However, only the interlayer optical field $(\bu_1+\bu_2-\bu_3-\bu_4)/4$ has a significant impact on electronic properties, the other 3 modes being at least an order of magnitude less in magnitude. 

\begin{figure}[hbp]
\centering
\includegraphics[width=0.66\textwidth]{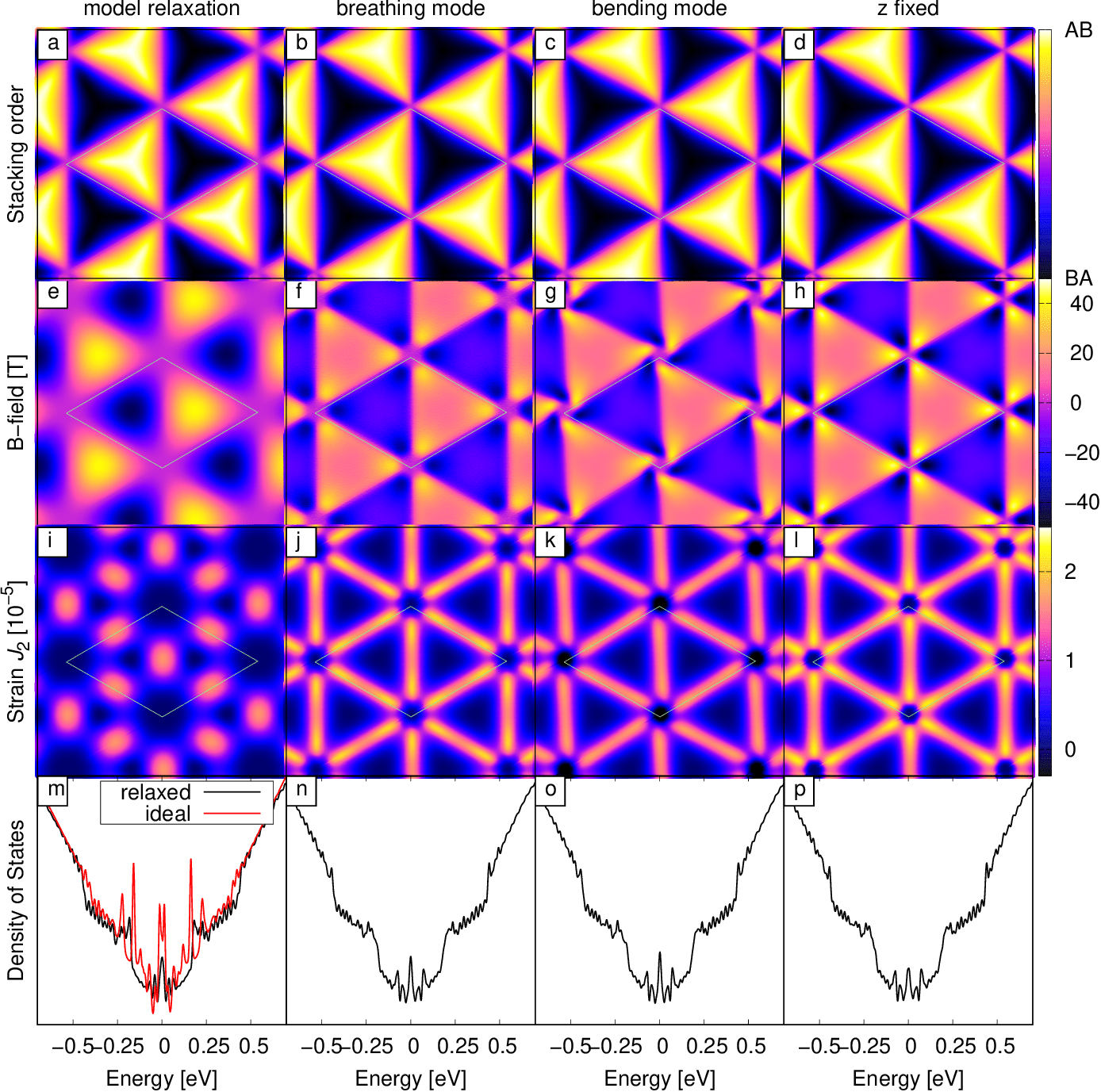}
\caption{\label{F:comp}
  Comparison of key structural and electronic features of the graphene twist
  bilayer ($\theta=0.51^\circ$, $V=0$) as obtained from a model relaxation
  field (MRF) and structural optimisation calculations. The first column
  is the MRF, with three structural types obtained by structure optimization in the subsequent 3 columns: the breathing mode,
  the bending mode, and an artificial simulation in which the atoms are
  constrained to lie in-plane. From the stacking order of the twist bilayer
  (a-d) we see that the MRF captures the gross feature of structural
  relaxation, namely, the significant increase in AB/BA stacking types, but that fine structure is smoothed out, compare the model pseudo-magnetic field (e) with (f-h), and the model von Mises strain (i) with (j-l). The impact
  on the density of states of the twist bilayer of relaxation is significant, with the reduction in zero mode peak height well captured by the model relaxation field.}
\end{figure}

\begin{figure}[hbp]
\centering
\includegraphics[width=0.35\textwidth]{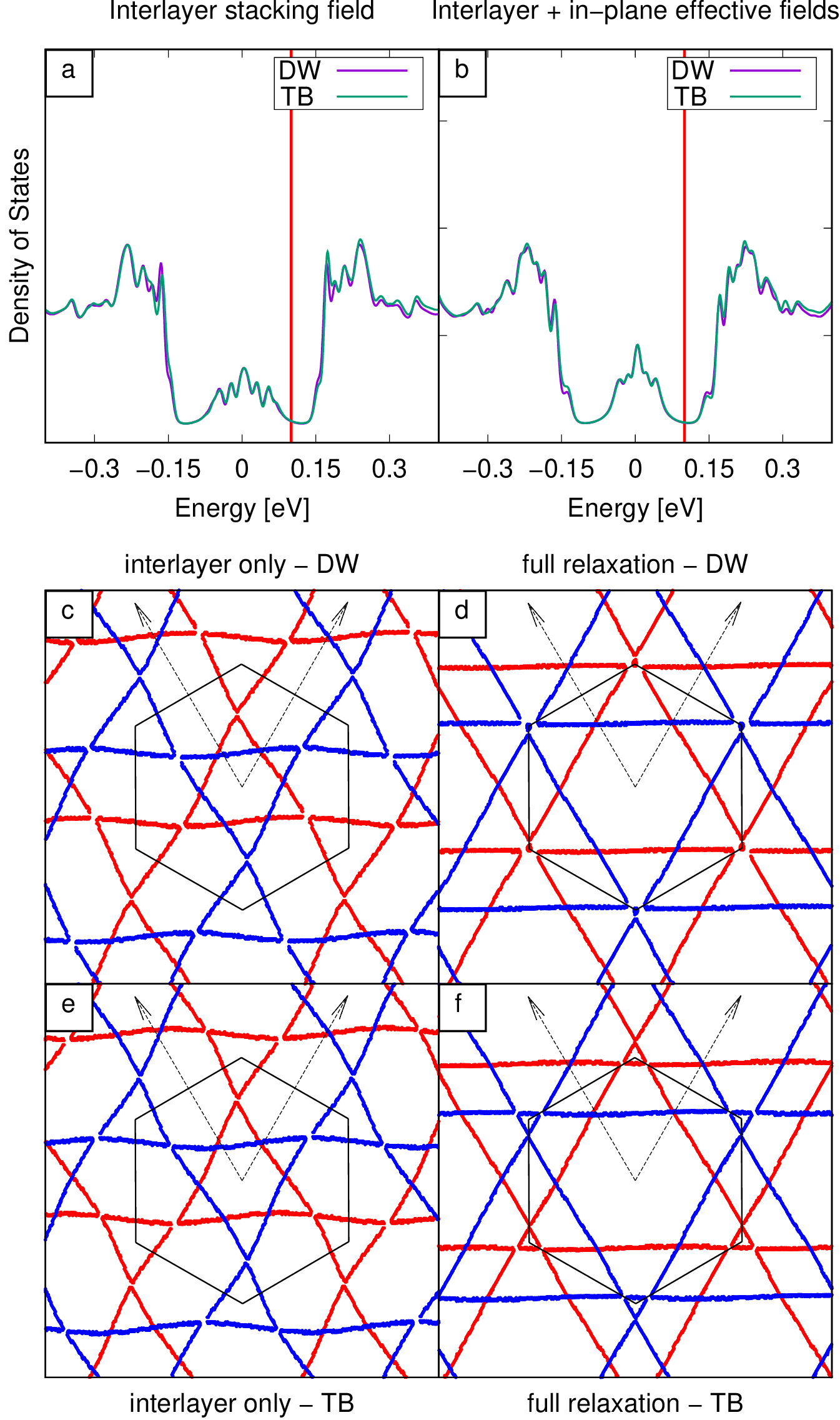}
\caption{\label{Rel}
  Shown are the density of states for a twist bilayer with
  $\theta=0.51^{\circ}$ and bias $U=\pm 300\,$meV with the single layer
  blocks treated as perfect single layer graphene (i.e., not including the
  impact of relaxation in the single layer blocks) at the Dirac-Weyl (DW)
  and tight-binding (TB) levels, panel (a), and with inclusion of relaxation
  pseudo-gauge and scalar terms, also at the DW and TB levels, panel (b). For
  both cases the interlayer interaction includes the effect of relaxation.
  As may be seen, while going from DW to TB generates somewhat greater
  particle-hole asymmetry, it does not substantially change the DOS. However,
  inclusion of the relaxation generated effective fields leads to a smoothing
  of the central Dirac peak and shifts the valley in energy. In panels (c-f)
  the same 4 levels of theory are illustrated for the Fermiology at 100\,meV.
  Again, while DW versus TB makes only a slight difference, inclusion of
  relaxation physics in the single layer blocks results in more substantial
  changes.}
\end{figure}

To analyze the impact of lattice relaxation on the twist bilayer we first consider the \emph{stacking order}. The interlayer interaction can be expressed as
\begin{equation}
 S(\br) = c_{AA}(\br)\begin{pmatrix} 0 & 1 \\ 1 & 0 \end{pmatrix}
 + c_{AB}(\br)\begin{pmatrix} 1 & 0 \\ 0 & 0 \end{pmatrix}
 + c_{BA}(\br)\begin{pmatrix} 0 & 0 \\ 0 & 1 \end{pmatrix}
\end{equation}
from which we can define the change in stacking order as $|c_{AB}(\br)|^2-|c_{BA}(\br)|^2$. The continuous modulation of stacking order that characterizes the ideal twist bilayer is replaced by trigonal domains of AB and BA stacking which, as can be seen in Figure\,\ref{F:comp} (a-d) is well captured by the relaxation model. Compare panel (a) which displays the change in stacking order due to the model relaxation field with panels (c-d) which exhibit the change due to three distinct relaxation fields that result from structural optimization: (b) the breathing mode (c) the bending mode and (d) relaxations confined to a plane.

The effective pseudo-magnetic field is given by the curl of the pseudo-gauge
\begin{equation}
\bA(\br) = (\varepsilon_{11}(\br) - \varepsilon_{22}(\br), 2\varepsilon_{12}(\br))
\end{equation}
with the components of the strain tensor given by \,$\varepsilon_{ij} = (\partial_i u_j + \partial_j u_i)/2$. The pseudo-magnetic field therefore involves third derivatives of the deformation field, and provides a closer look at the impact of lattice relaxation. The strength of the pseudo-magnetic field is significant, Figure\,\ref{F:comp}(e-h), with its magnitude well captured by the relaxation model. However, both the partial dislocations and the nodes of the dislocation network are seen to be much smoother in model relaxation than in structure optimization, with the fine structure at the nodes of the network not captured. This can be seen also in the von Mises strain, i.e.~the second invariant
\begin{equation}
J_2 = (\varepsilon_{11} - \varepsilon_{22})^2 +
      (\varepsilon_{22} - \varepsilon_{33})^2 +
      (\varepsilon_{33} - \varepsilon_{11})^2 -
      2\,\varepsilon_{12}\varepsilon_{21} -
      2\,\varepsilon_{23}\varepsilon_{32} -
      2\,\varepsilon_{31}\varepsilon_{13}
\end{equation}
see Figure\,\ref{F:comp}(i-l). This is a measure of the distortion energy of a deformation and, as the partial dislocations are pure shear, is a good measure of the location of the dislocation lines. As can be seen by comparing panel (i), the von Mises strain from the model relaxation, with panels (j-l), the formation of dislocations is only incompletely captured as compared to the structural optimization  calculations.

The electronic consequences of the lattice reconstruction are significant, see Figure\,\ref{F:comp}m. Lattice relaxation leads to pronounced reduction in the Dirac point zero mode with the reduction in peak height well captured by the relaxation model, although differences exist in fine structure of the spectrum, which differs also between different relaxation types, see Figure\,\ref{F:comp}(n-p).

\clearpage
\subsection{Impact of tight-binding versus Dirac-Weyl, and relaxation corrections to the layer diagonal blocks}

\clearpage
\subsection{Overview of local density of states as a function of Fermi energy}

Close to the Dirac point charge localization occurs on AA regions of the lattice, however under an applied bias the valley region, between 40-110~meV, exhibits a very different physics. There are three distinct structures seen in the local density $\rho(\br,E)$: (i) expulsion of charge from the AA regions, (ii) localization on the partial dislocations, and (iii) a bright ``halo'' of localization surrounding the AA regions. These features can be seen in a recent STM investigation of the twist bilayer under bias \cite{Huang2018}.

\begin{figure}[!h]
\centering
\includegraphics[width=0.56\textwidth]{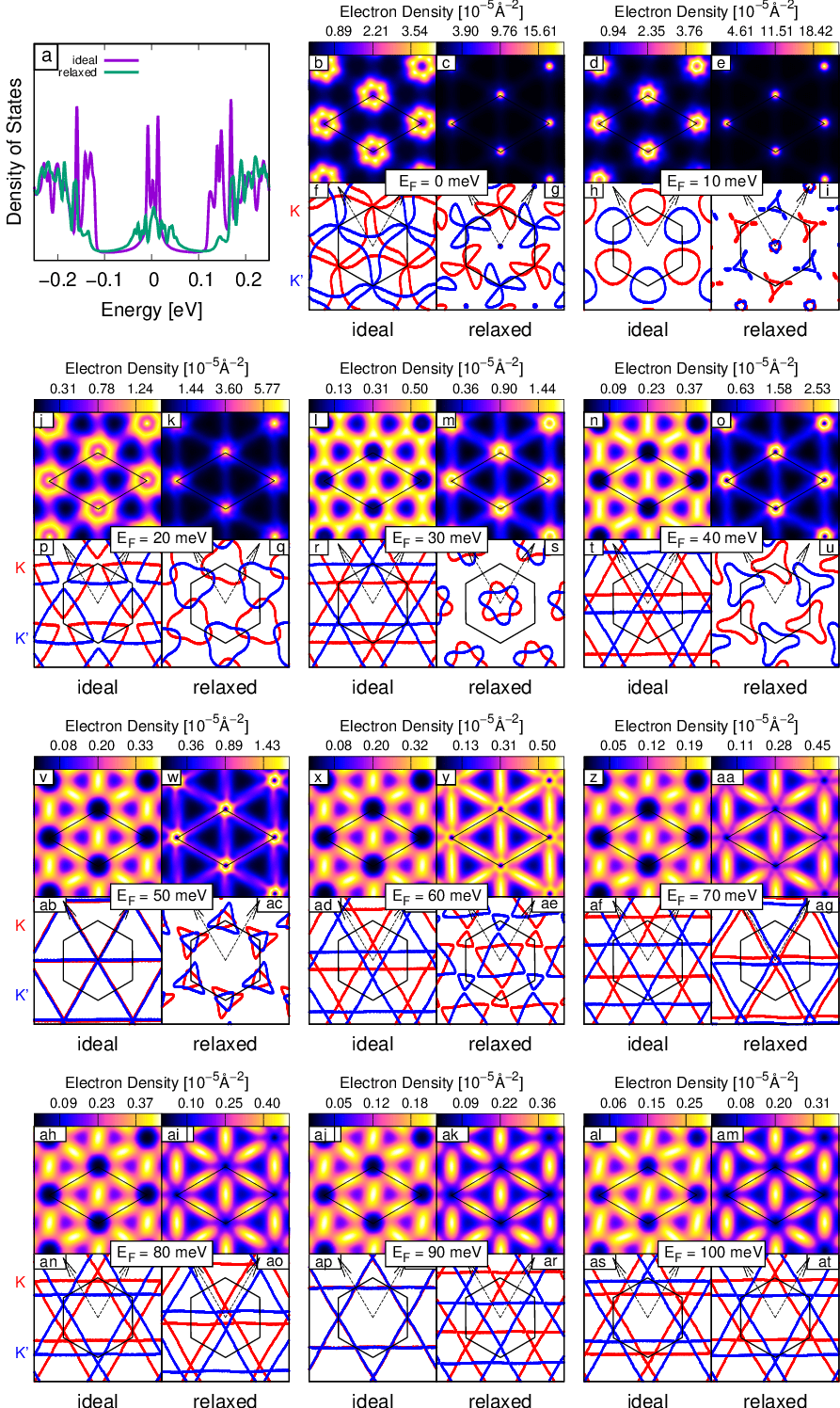}
\caption{
  Fermiology and density of states for the ideal twist bilayer and the
  reconstructed dislocation network of a bilayer with $\theta=0.51^{\circ}$
  and bias $U=\pm 300\,$meV, for an energy range spanning 0 to 100\,meV.}
\end{figure}

\clearpage
\subsection{Nesting phase diagram for the ideal twist bilayer}

\begin{figure}[!h]
\centering
\includegraphics[width=0.60\textwidth]{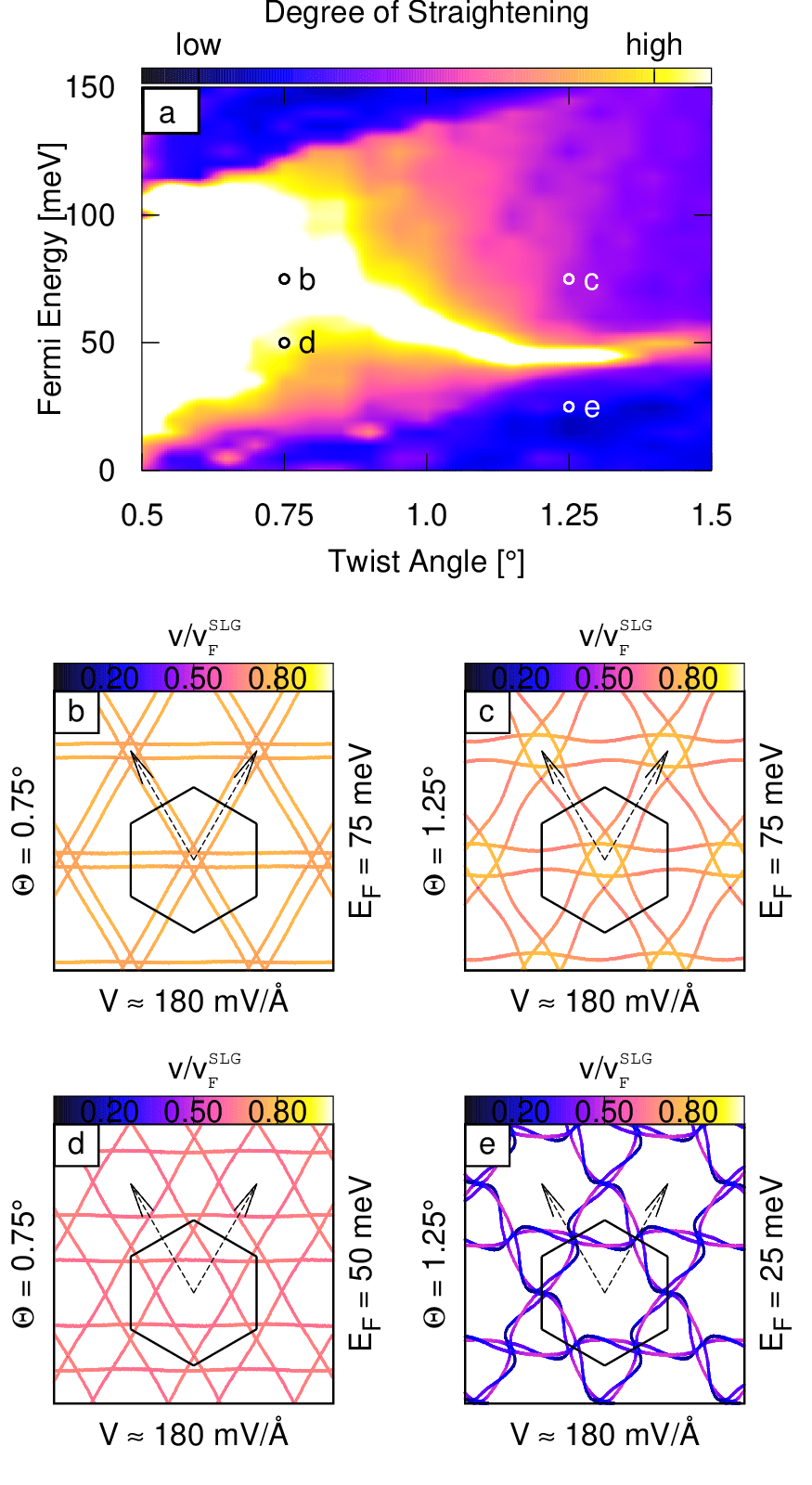}
\caption{
  Nesting phase diagram for the ideal twist bilayer, for bias $U=\pm 300$\,meV.
  This can be compared with the nesting phase diagram for the reconstructed
  partial dislocation network, see Fig.\,2 of the main text.}
\end{figure}

\clearpage
\subsection{Robustness of nesting with respect to parameters of a simplified model}

\begin{figure}[!h]
\centering
\includegraphics[width=0.80\textwidth]{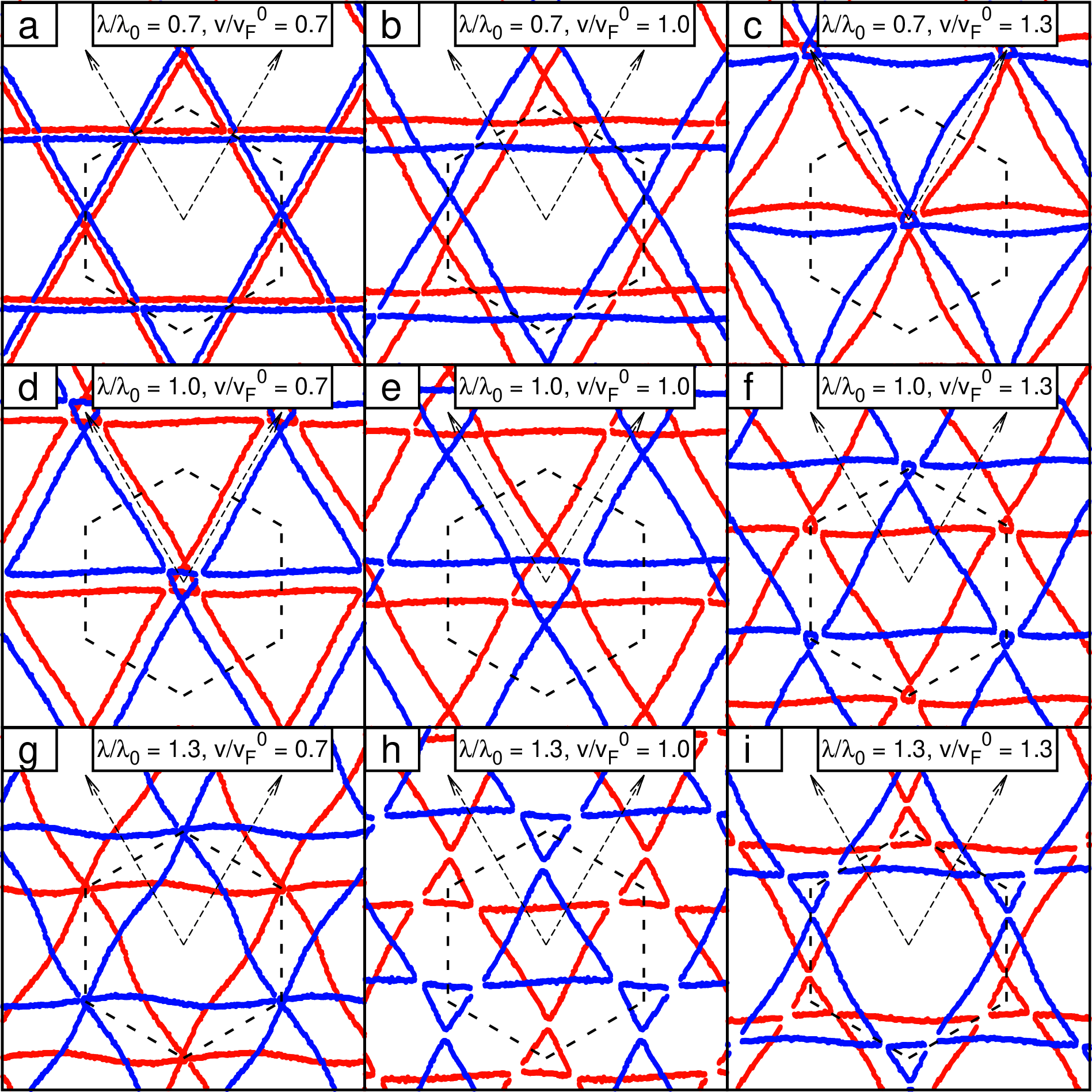}
\caption{
  In a simplified model of the twist bilayer one has a two parameter theory:
  the Fermi velocity ($v_F$) of the Dirac-Weyl approximation to the single
  layer constituents and (ii) the coupling strength ($\lambda$) of the
  interlayer moir\'e field. The nesting effect (which is well captured by
  this simple model) is robust against these two parameters, as can be seen
  in panels (a-i) above in which a substantial range of values of $v/v_F^0$
  and $\lambda/\lambda_0$ are examined (with $v_F^0$ and $\lambda_0$ the
  equilibrium values). However, the nesting vector changes with variation
  of these parameters.}
\end{figure}


\end{document}